\documentclass[%
amsmath,amssymb,
 aps,
pra,twocolumn,showpacs
]{revtex4-1}

\usepackage{graphicx}% Include figure files
\usepackage{dcolumn}% Align table columns on decimal point
\usepackage{bm}% bold math
\usepackage{hyperref}% add hypertext capabilities
\usepackage{color}\usepackage[normalem]{ulem}
\usepackage{epstopdf}

\usepackage{units}

\begin{document}

%\preprint{APS/123-QED}

\title{Monitoring squeezed collective modes of a 1D Bose gas after an interaction quench using density ripples analysis}

\author{Max Schemmer}
%%  \affiliation{Laboratoire Charles Fabry, Institut d’Optique, CNRS, Universit\'{e} Paris Sud 11,
%% 2 Avenue Augustin Fresnel, F-91127 Palaiseau Cedex, France}
\author{Aisling Johnson}
\altaffiliation[current address: ]{Vienna Center for Quantum Science and Technology, TU Wien-Atominstitut, Stadionallee 2, 1020 Vienna, Austria.}%
\author{Isabelle Bouchoule}%
 \email{isabelle.bouchoule@institutoptique.fr}
\affiliation{Laboratoire Charles Fabry, Institut d’Optique, CNRS, Universit\'e Paris Sud 11,
2 Avenue Augustin Fresnel, F-91127 Palaiseau Cedex, France}%

%\collaboration{}%\noaffiliation

\date{\today}% It is always \today, today,

\begin{abstract}
 We  investigate the out-of-equilibrium dynamics 
 following a sudden quench of the interaction strength, 
in a one-dimensional quasi-condensate trapped at the surface of an atom chip.
Within a linearized approximation,   the system is described by independent collective modes and  
the quench squeezes the phase space 
distribution of each mode, 
leading to a subsequent breathing of each quadrature.
We show that 
the collective modes are resolved by the  power spectrum of 
density ripples which appear after a 
short time of flight.
This allows us to experimentally probe the expected breathing phenomenon. 
Our results are in good agreement with theoretical predictions which take  the 
longitudinal harmonic confinement into account. 
\end{abstract}

\pacs{03.75.Hh, 67.10.Ba}

\maketitle

\section{Introduction}
 The out-of-equilibrium dynamics of isolated quantum many-body systems
is a field attracting a lot of interest~\cite{polkovnikov_colloquium:_2011}, triggered 
in part by progress in cold atom experiments.
A particular focus has been devoted to the case of sudden quenches where 
the system is brought out-of-equilibrium by a sudden change of a Hamiltonian 
parameter,  and in particular the case of an interaction
quench, 
both  theoretically~\footnote{See~\cite{mitra_quantum_2017} and references
  therein} and
experimentally~\cite{trotzky_probing_2012,cheneau_light-cone-like_2012,hung_cosmology_2013,langen_double_2017,jaskula_acoustic_2012}. 
Whether and how the system relaxes towards an equilibrium state is 
the subject of intense theoretical work.
The role of integrability, 
not completely elucidated, 
is the focus of many studies. 
Within this context,
the case of a 1D Bose gas 
with contact repulsive interactions, described by the integrable
Lieb-Liniger model,  is a prime theoretical candidate to uncover
  the underlying physics,
  studied in \textit{e.g} ~\cite{de_nardis_solution_2014,calabrese_interaction_2014,cazalilla_quantum_2016,swislocki_quantum_2016}.

 This paper constitutes the experimental study of the 
out-of-equilibrium dynamics  following a sudden quench of the 
interaction strength in a 1D Bose gas with repulsive interactions.
Within a linearized approximation, the evolution following a splitting
of a 1D Bose gas in two copies, studied in~\cite{langen_double_2017},
can be interpreted as an interaction quench in an effective 1D Bose gas.
Investigating the first-order correlation function, 
the authors observed an apparent thermalization, taking the form of a 
light cone effect. This observation may however conceal underlying 
non-equilibrium dynamics,  
as revealed recently by the observation
of recurrences in a similar experiment~\cite{rauer_recurrences_2017}.
Finding appropriate observables revealing these 
dynamics is thus a key point for investigating out-of-equilibrium 
phenomena.
 In this paper, by investigating the density ripples
appearing after short time of flight, the behavior of collective modes is probed, 
rather than a global quantity such as the first-order correlation function,
allowing for a better understanding
of the physics at play after an interaction quench.
 The dynamics is revealed
by the  oscillatory behavior of each component of the
density ripples power spectrum, observed for times that go beyond
the apparent thermalization time seen on the first order correlation
function.
We show that these oscillatory dynamics are the signature of
squeezed collective modes: for each collective mode, the quench produces a
squeezed phase space distribution, leading to a subsequent oscillation
of the width of its quadratures --- a \textit{breathing behavior}.
As well as improving the understanding of the effect of an interaction quench,
this work constitutes an observation of squeezed collective modes,
a result interesting on its own.

\begin{figure}
\centerline{\includegraphics[width=1\columnwidth]{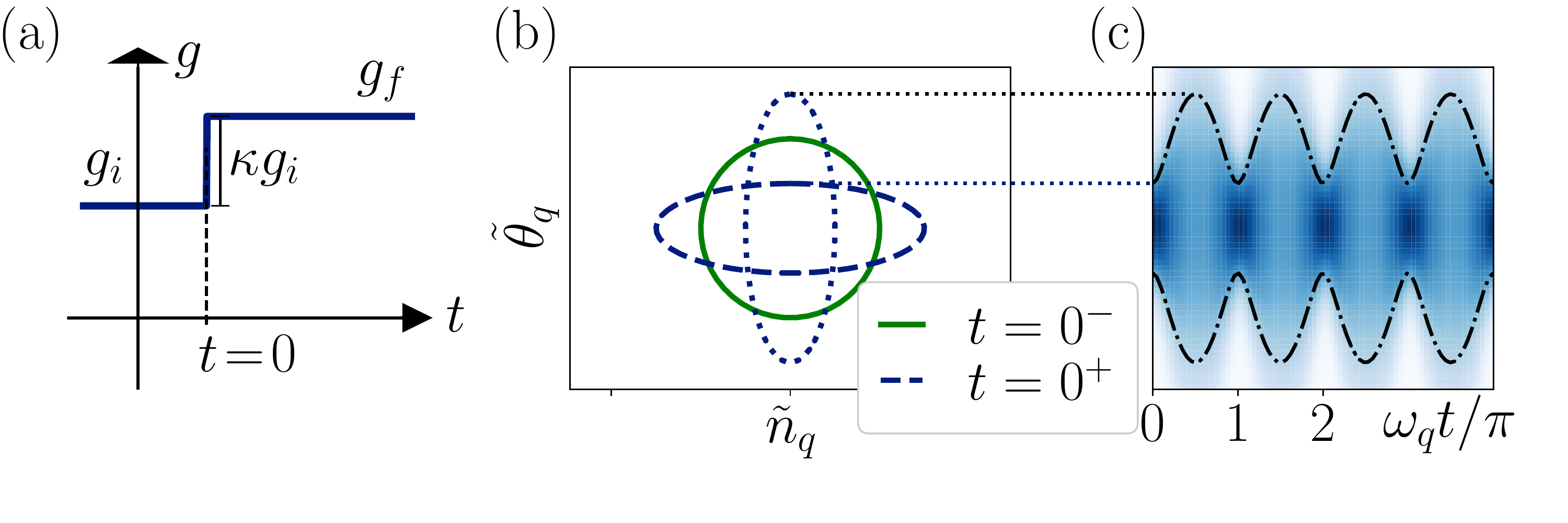}}
\caption{Squeezing of each collective mode after an interaction  strength quench from
 $g_i$ to $g_f$.
The Gaussian  phase  space distributions before the quench ($t=0^-$), 
just after the quench ($t=0^+$) and after an evolution 
time $\pi/\left(2 \omega_q\right)$ (dashed ellipse) are represented 
in (b), where lines 
correspond to a given probability density  (here we chose $\kappa=3$).
The subsequent 
breathing is seen in (c), where the time evolution of the phase distribution is shown
in color plot.
%% Squeezing of each collective mode after an interaction strength quench.
%% The Gaussian space distributions before the quench (solid circle), 
%% just after the quench (dashed ellipse) and after an evolution 
%% time $\pi/\left(2 \omega_q\right)$ (dotted ellipse) are represented 
%% in (b), where lines 
%% correspond to a given probability density. The subsequent 
%% breathing is seen in (c), where the time evolution of the phase distribution is shown.
}
\label{fig.phasespace}
\end{figure}

\section{The interaction quench within the linearized approach}
The physics at play can be understood by considering a 1D homogeneous Bose 
gas, 
of length $L$, temperature $T$ and density $n_0$,
with particles of mass $m$ interacting with a two-body repulsive contact 
interaction $g \delta(z)$, where $z$ is the distance between the two particles. 
At $t = 0$, $g$ is suddenly changed from $g_i$ to $g_f = (1+\kappa) g_i$,
where $\kappa$ is the quench strength. 
  While the complete treatment of an interaction quench
  is tremendously difficult
  the problem is greatly simplified if one can rely on a linearized approach, as
  presented below.
Within the quasi-condensate regime, 
density fluctuations are strongly reduced ($|\delta {n}(z)| \ll n_0 $) 
and phase fluctuations occur on large length scales, such that the 
Hamiltonian of the system  can be
diagonalized using the phase-density
representation of the field operator
$ \Psi(z) = \sqrt{n_0 + \delta {n}(z)} \exp(i{\theta}(z))$
and  the Bogoliubov procedure~\cite{mora_extension_2003}. The obtained 
linearized modes correspond to Fourier modes. For each wave-vector $q$, 
the dynamics is governed by the harmonic oscillator Hamiltonian~\cite{schemmer_monte_2017}
\begin{equation}\label{Bogo}
H_q = A_q {n}_{q}^2 + B_q {\theta}_q^2  = \hbar \omega_q \left( {\tilde{n}_q^2}/{2} + {\tilde{\theta}_q^2}/{2}  \right)
\end{equation}
where  the canonically conjugated hermitian operators ${n}_{q}$ and ${\theta}_q$ are the Fourier 
components~\footnote{For each positive $q$ value, one has 2 Fourier components:
\unexpanded{$\hat{n}_{q,c}=\sqrt{2/L}\int dz n(z)\cos(qz)$}  and 
\unexpanded{$\hat{n}_{q,s}=\sqrt{2/L}\int dz n(z)\sin(qz)$}, with similar expressions for $\theta$.  
We omit the subscript $c$ or $s$ in the text for simplicity.} 
of
$\delta n$ and $\theta$ and where the reduced variables are defined by 
$\tilde{n}_q = n_q ( A_q/B_q)^{1/4}$ and $\tilde{\theta}_q = \theta_q (B_q /A_q)^{1/4}$.
For wavevectors $q$ much smaller than the inverse healing length $\xi^{-1} = \sqrt{m g n_0}/\hbar$,  the excitations are of hydrodynamic nature~\footnote{For quasi-1D gases the hydrodynamic condition is replaced by $\omega_q \ll \omega_{\perp}$.}. 
Their frequency is $\omega_q=cq$, where the speed of sound is
$c=\sqrt{n_0\partial_n\mu/m}$, and the Hamiltonian's coefficients are 
$B_q=\hbar^2 q^2 n_0/(2 m)$ and  $A_q= mc^2/(2n_0)$.
Here $\mu(n)$ is the equation of state of the gas relating the chemical potential $\mu$ 
to the linear density, which reduces to  $\mu=gn$ for
pure 1D quasi-condensate.
For a given $q$, the dynamics of the quenched harmonic oscillator is represented in Fig.~(\ref{fig.phasespace}). 
Before the quench 
%the system is in thermal equilibrium and 
the 
phase space distribution is the one of a thermal state: an isotropic Gaussian
 in the ($\tilde\theta_q, \tilde{n}_q$)-plane.
The quench affects $A_q$ 
while $\theta_q$ and $n_q$ do not have time to change. The variances thus become 
$\langle \tilde{\theta}_q^2 \rangle_{t=0^+} = \langle \tilde{\theta}_q^2 \rangle_{t=0^-}/(1+\kappa)^{1/2}$ 
 and
$\langle \tilde{\delta n_q}^2 \rangle_{t=0^+} = \langle \tilde{\delta n_q}^2 \rangle_{t=0^-}(1+\kappa)^{1/2}$\footnote{The phase space area is preserved, one quadrature being squeezed, while the other 
is anti-squeezed.}.
The subsequent evolution is a rotation in phase space at a frequency  
$\omega_q$ leading to a breathing of each quadrature. In particular
\begin{equation}
\langle \theta_q^2\rangle = \langle \theta_q^2\rangle_i(1+\kappa\sin^2(cqt)),
\label{eq.osctheta2}
\end{equation}
where  the  initial value 
$\langle \theta_q^2\rangle_i$ is the thermal prediction
$\langle \theta_q^2\rangle= m k_B T/(\hbar^2n_0 q^2)$~\footnote{For the 
$q$ values considered, $\hbar \omega_q \ll k_B T$ and the Raighley-Jeans approximation holds.}.

Probing the non equilibrium dynamics following a quench is not straightforward, 
especially concerning the choice of the observable.  Since density fluctuations are very
small within the
 quasi-condensate regime, it is more advantageous to probe the phase
 fluctuations~\footnote{In~\cite{hung_cosmology_2013}, the evolution of density fluctuations
   has however been investigated for a 2D gas.}. 
 One way is to investigate the one-body correlation function
 $g_1(z) =\langle\hat{\Psi}^{\dagger}(z) \hat{\Psi}(0)\rangle$, which,
 for $z\gg \xi$ and in the quasi-condensate regime, writes
 $g_1(z) \simeq n_0 e^{-\langle (\theta(z)-\theta(0))^2\rangle /2}$~\cite{mora_extension_2003}.
 However since phase
 fluctuations are large in a quasi-condensate, the exponential cannot
 be linearized and  $g_1(z)$  mixes contributions from all Bogoliubov
 modes~\footnote{Isolating the contribution of individual modes to the
   function $g_1(z)$ requires looking at the Fourier transform of
   $\ln (g_1(z))$, which requires large detection
   dynamics.}, preventing the observation of  the
squeezed dynamics. In
 fact, the linearized model above predicts the light-cone effect on
 the $g_1$ function:   $g_1(z)$ changes from its initial
 exponential decay $\exp(-|z|/l_c^i)$, where $l_c^i = 2 \hbar^2 n_0/( m k_B T)$, to an exponential decay with
 a new correlation length $l_c^f =2 l_c^i/ (\kappa+2)$ for
 $z< 2 c t$.  The breathing of each squeezed Bogoliubov
 mode is not transparent here. 
Moreover, 
for times larger
than a few $t^{g_1}_{\rm th}= l_c^f /c$, the $g_1$ 
function essentially reaches the form expected for a thermal state at a
temperature $T_f =T(\kappa+2)/2$, and the ongoing dynamics
is hidden.
  In this paper we use the density ripples analysis to
 reveal the non equilibrium dynamics of the gas by probing the
 breathing of each mode.

\section{Resolving Bogoliubov modes with density ripples}
Density ripples appear  after switching the interactions off and waiting for a free evolution time $t_f$ (time-of-flight), during which phase fluctuations transform into 
density fluctuations~\cite{imambekov_density_2009,dettmer_observation_2001,manz_two-point_2010,rauer_cooling_2016}.
%Density ripples appear  after switching the interactions off and waiting for a free evolution time $t_f$ (time-of-flight), during which phase fluctuations transform into 
%density fluctuations~\cite{imambekov_density_2009}.
%The analysis of these density ripples has been used as 
%thermometry~\cite{dettmer_observation_2001,manz_two-point_2010}, 
%and to investigate the cooling mecanism~\cite{rauer_cooling_2016}.
Consider the power spectrum of density ripples
$\langle |\rho(q)|^2\rangle$, where 
$\rho(q)=(1/\sqrt{L})\int d z (\langle  n(z,t_{f}) -n_0) e^{ i q z}$.
Propagating the field operator during the time of flight  and
assuming translational invariance we 
obtain~\footnote{For consistency we rederive this expression
(first established in~\cite{imambekov_density_2009})
see Appendix~\ref{SM_Density_ripples_hom},\ref{SM_Density_ripples_LDA}.}
\begin{equation} 
\langle |\rho_{n_0}(q)|^2\rangle= \int dx e^{-iqx}(f(q,x) -n_0^2),
\label{eq.rhoq2}
\end{equation}
where 
\begin{equation} 
%\begin{array}{lll}
%f(q,x)&=& \langle \psi^+(0)\psi(-\hbar qt_f/m)\psi^+(x-\hbar qt_f/m)\psi(x)\rangle\\ &\simeq&
%n_0^2\langle e^{i \left[ \theta(0)-\theta(-\hbar qt/m)+\theta(x-\hbar qt/m)-\theta(x) \right] }\rangle,
%\end{array}
f(q,x)\simeq
n_0^2\langle e^{i \left[ \theta(0)-\theta(-\hbar qt_f/m)+\theta(x-\hbar qt_f/m)-\theta(x) \right] }\rangle,
\label{eq.jrho}
\end{equation}
averages in Eq.~(\ref{eq.jrho}) are taken before the time of flight.
The function $f$ involves only 
pairs of points separated by $\hbar q t_{f}/m$. For small wave 
vectors $q\hbar t_{f}/m\ll l_c$,
the phase difference between those points is small and one can expand the exponential. 
To lowest order, assuming uncorrelated distributions for each mode $q$ and vanishing mean values, 
we then find 
\begin{equation}
\langle |\rho_{n_0}(q)|^2\rangle = 4 n_0^2 \langle \theta_q^2\rangle \sin^2\left({\hbar q^2 t_f}/{(2m)}\right),
\label{eq.drasymp}
\end{equation}
showing that, for low lying $q$, the density ripples spectrum directly resolves  the phase 
quadrature $\langle \theta_q^2\rangle$ of individual Bogoliubov 
modes~\footnote{In Eq.~(\ref{eq.drasymp}),
\unexpanded{$\langle \theta_q^2 \rangle =(\langle \theta_{q,c}^2 \rangle+\langle \theta_{q,s}^2 \rangle)/2$} 
where $\theta_{q,c}$ and $\theta_{q,s}$ are the cosine and sine Fourier components, which 
fulfill \unexpanded{$\langle \theta_{q,c}^2\rangle=\langle \theta_{q,s}^2\rangle$} 
for translationally invariant systems.
}.
The proportionality between $\langle |\rho_{n_0}(q)|^2\rangle$ 
and $\langle \theta_q^2\rangle$ implies that $\langle |\rho_{n_0}(q)|^2\rangle$ 
oscillates according to Eq.~(\ref{eq.osctheta2}) when varying the time $t$ after the quench. 
Density ripples are thus an ideal tool to investigate the quench dynamics.
Note that, in the following we are interested, for each wave vector $q$,
  in the evolution of $\langle |\rho_{n_0}(q)|^2\rangle$ with
  the evolution time $t$, such that the proportionality
  factor $4 n_0^2  \sin^2\left({\hbar q^2 t_f}/{(2m)}\right)$
  is irrelevant for our data analyis.

In typical experiments, atoms are confined by a smooth potential $V(z)$.
% generally harmonic
%, 
%complicating the picture.
%However, 
For weak enough  confinement  and for wavelengths 
much smaller than the system's size, one can however use 
the above results  for homogeneous systems 
within  
a local density 
approximation (LDA)~\footnote{Validity of LDA is established in Appendix~\ref{sec.beyondLDA}}. 
 Then
$\tilde{\rho}(q)=\int dz \delta n (z,t_f)e^{iqz}$ fulfills 
$\tilde{\rho}(q)\simeq\int dz \langle |\rho_{n_0(z)}(q)|^2\rangle$  
where 
$n_0(z)$ is the
density profile, which can  
itself be evaluated within the LDA using 
the gas equation of state and the local chemical potential $\mu(z)=\mu_0
-V(z)$. Injecting  Eq.~(\ref{eq.osctheta2}) and Eq.~(\ref{eq.drasymp})
into the LDA integral, we find 
\begin{equation}
\langle |\tilde{\rho}(q)|^2\rangle/\langle |\tilde{\rho}(q)|^2\rangle_i 
=1+\kappa{\cal F}(cqt),
\label{eq.calF}
\end{equation}
where $c$ is the speed of sound after the quench 
evaluated at the trap center and
${\cal F}$ 
only depends on the shape of $V(z)$.
 For a box-like potential, one recovers previous results and
${\cal F}(\tau)=\sin^2(\tau)$.
The  expression 
of ${\cal F}$  is given in~Appendix~\ref{SM_evol_LDA} in the case of a harmonic potential:
The oscillatory behavior is preserved, although
the spread in frequencies originating from the 
inhomogeneity in $n_0$ introduces
damping, which is 
a pure dephasing effect.
%% where $c$ is the speed of sound after the quench
%% evaluated at the trap center and ${\cal F}$ 
%% only depends on the shape of $V(z)$. 
%% The  
%% expression 
%% of ${\cal F}$ 
%% for a harmonic potential is given in Appendix~\ref{SM_evol_LDA}.
%% %Since the central part of the cloud gives the dominant contribution, 
%% \unexpanded{$\langle |\tilde{\rho}(q)|^2\rangle$} 
%% still presents an oscillatory behavior as 
%% a function of $t$, ${\cal F}(\tau)$ being close to $\sin^2(\tau)$.
%%  The spread in frequencies originating from the 
%% inhomogeneity in $n_0$ is 
%% however responsible for a damping, which is 
%% a pure dephasing effect.

\section{Experimental realization}
The experiment uses an atom-chip set
up~\footnote{The experiment is described in more detail in~\cite{jacqmin_momentum_2012}.} 
where  $^{87}$Rb atoms are magnetically confined using
current-carrying micro-wires.  The transverse confinement, acting in a
vertical plane, is provided
by  three parallel wires carrying AC-current modulated at
  400~kHz, which renders the magnetic potential insensitive to wire
  imperfections and, allows for independent control of the transverse
  and longitudinal confinements.  We perform radio
frequency (RF) forced evaporative cooling until we reach the desired
temperature. We then increase the RF frequency by
\unit[60]{kHz}, providing a shield for energetic three-body collision
residues and wait during \unit[150]{ms} relaxation time%% ~\footnote{ We are
%%   aware that this initial state does not necessarily represent a
%%   thermal state~\cite{johnson_long-lived_2017}.  
%% However, the density ripples analysis only probes
%%   phonons, and their distribution is consistent with thermal
%%   equilibrium.}
.  The clouds contain a few thousand atoms, in a trap
with a transverse frequency $\omega_{\perp}/ 2\pi = $ \unit[1.5]{} or \unit[3.1]{kHz},
depending on the data set,
%transverse frequency $\omega_{\perp}/ 2\pi = $ \unit[1.5 - 3.1]{kHz}
and a longitudinal frequency $\omega_{\parallel}/ 2\pi =
$ \unit[8.5]{Hz}. The samples are quasi-1D, the temperature and 
chemical potential satisfying $\mu,k_B T < \hbar \omega_{\perp}$.
The temperature is low enough so that the gas typically lies well
within the quasi-condensate regime~\cite{kheruntsyan_pair_2003}.  
The equation of state is well
described by $\mu = \hbar \omega_{\perp} (\sqrt{1+4 n a} - 1)$, where
$a=5.3~$nm is the 3D scattering length~\cite{fuchs_hydrodynamic_2003}. 
While, for $na \ll 1$, one
recovers the pure 1D expression $\mu = g n$, where $g = 2 \hbar
\omega_{\perp} a$, this equation of state takes the 
broadening of the transverse size at larger $n a$ into account. 
The longitudinal density profile, well described by the LDA, 
extends over twice the Thomas-Fermi radius
$R_{TF}=\sqrt{2\mu_0/m}/\omega_{\parallel}$.  The speed of sound derived from the
equation of state is 
$c=c_{\rm{1D}}/(1+4na)^{1/4}$
where $c_{\rm{1D}}=\sqrt{2\hbar\omega_\perp na/m}$ is the pure 1D expression.
For data presented in this paper, $c/c_{\rm{1D}}$ is close to 0.7.
Since the effective interaction strength is proportional to $c^2$,
it is proportional to $\omega_{\perp}$. 

\begin{figure}[bp]
  %\centerline{%\raisebox{0.9cm}
  \vspace{-0.5cm}
  \hspace{1.8cm}
  \includegraphics[width=0.7\linewidth]{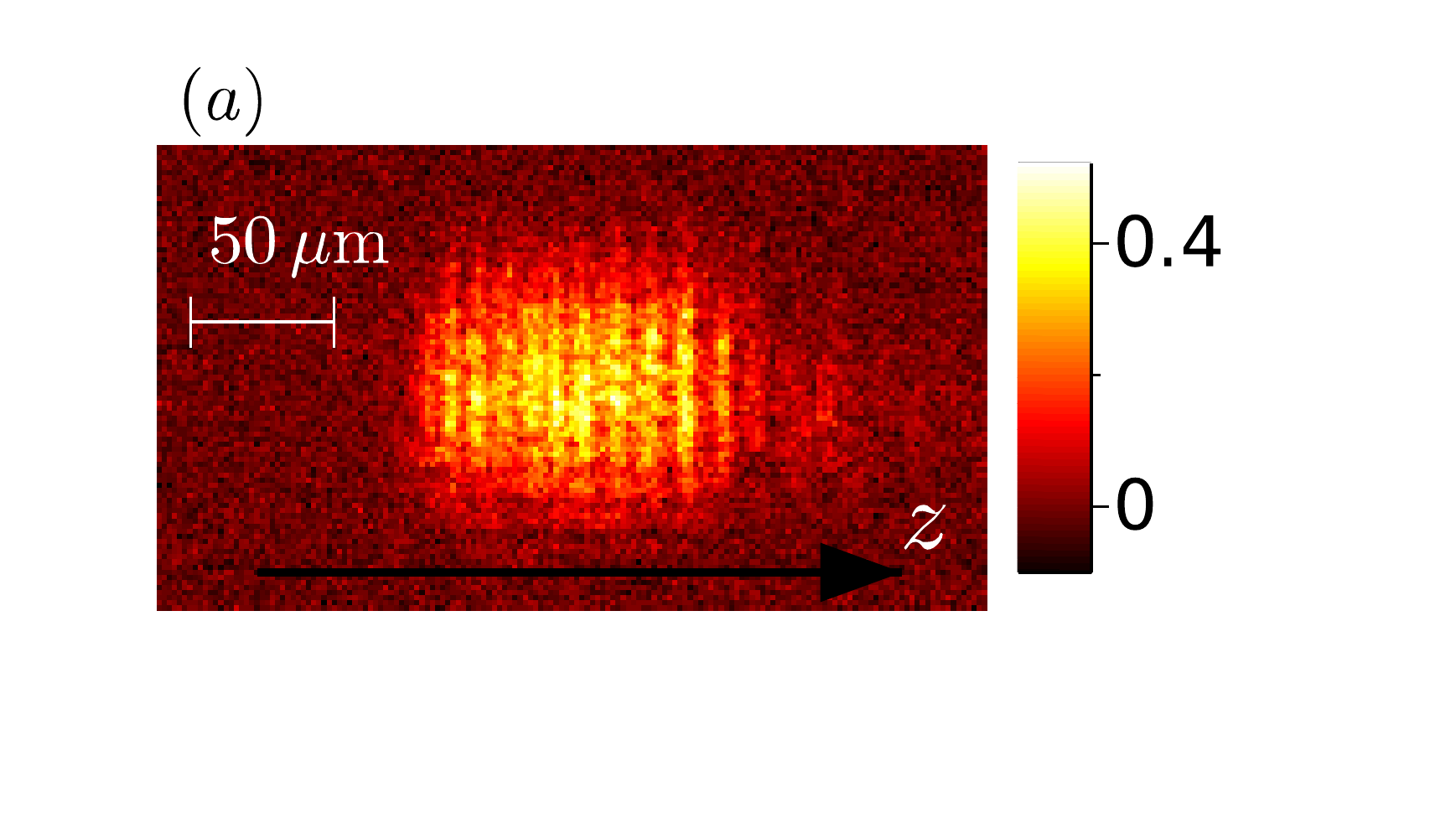}
%\includegraphics[width=0.45\linewidth,viewport=30 1 425 275,clip]{densityripples-image-article.pdf}}
%}
  %\hspace{-1cm}
  \vspace{-0.5cm}
  
\centerline{
\includegraphics[width=1\linewidth]{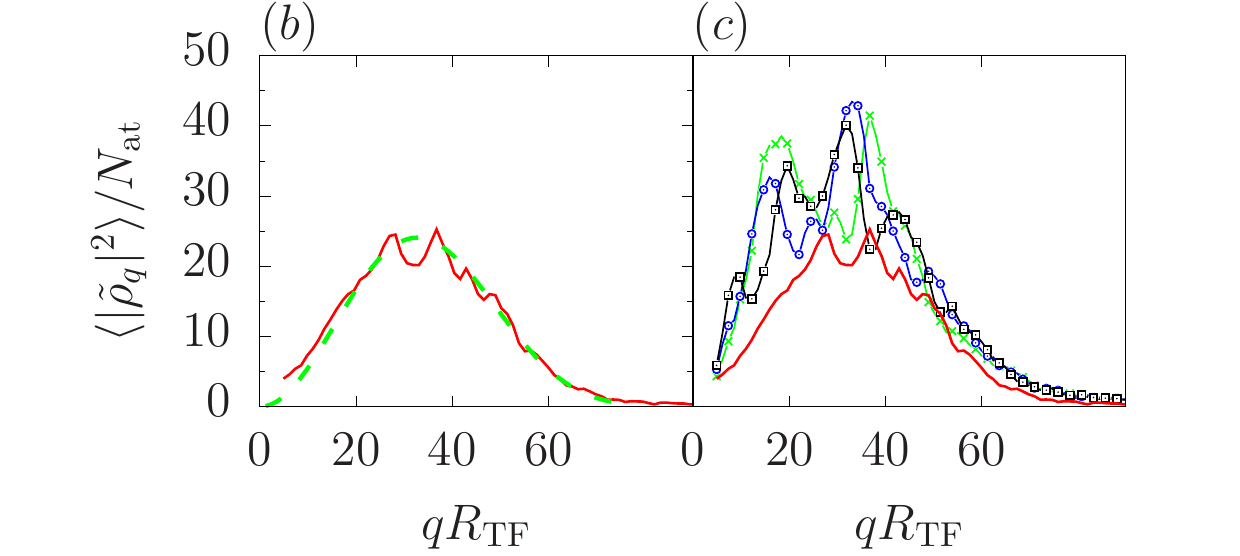}}
\caption{
Density ripples analysis (color online). 
$(a)$ Typical absorption image (optical density shown) 
taken after a time-of-flight  $t_f=8~$ms. 
$(b)$ Power spectrum of density ripples, obtained by averaging over about 50 images,
for a cloud at thermal equilibrium 
containing 16000 atoms confined in a trap with frequencies $\omega_z/(2\pi)=8.5~$Hz and $\omega_\perp/(2\pi)=1.5~$kHz, 
yielding a Thomas-Fermi radius $R_{TF} = 75~\mu$m.
The dashed (green) line is a theoretical fit (see text), yielding a temperature $T=55~$nK 
and an optical resolution $\sigma=2.9~\mu$m.
$(c)$ Power spectra after a quench of strength $\kappa=2$, at times $t=2.1~$ms (crosses, green), $t=2.6~$ms
(circles, blue) and 
$t=4.6~$ms (squares, black), the solid (red) curve being the initial power spectrum. 
}
\label{fig.densityripples}
\end{figure}

The interaction strength quench is performed by
ramping the transverse trapping frequency
$\omega_{\perp}$ from its initial value $\omega_{\perp,i}$ to its final value 
$\omega_{\perp,f}=(1+\kappa) \omega_{\perp,i}$
within a time $t_{r}$, typically of the
order of \unit[1]{ms}. This time is long enough 
for the transverse motion of the atoms to follow adiabatically but
short enough so that the quench can be considered as 
almost instantaneous with respect to the 
probed longitudinal excitations (see Appendix~\ref{SM_ramp_time}).
 We simultaneously 
multiply $\omega_\parallel$ by $\sqrt{1+\kappa}$,
to avoid modification of 
the mean profile
and 
of the Bogoliubov wavefunctions(see Appendix~\ref{sec.beyondLDA}).

 In order to probe density ripples, we release the atoms from the trap and let
them fall under gravity for a time $t_{f} = 8$~ms before taking an absorption image.
The transverse expansion, occurring on a time scale of $1/\omega_\perp$, ensures
the effective instantaneous switching off of the interactions with respect to 
the probed  longitudinal excitations.
The  density ripples produced by the phase fluctuations present before the free fall
are  visible in each individual image, as seen in Fig.~(\ref{fig.densityripples})$(a)$. 
From the image,  we record the longitudinal density profile $\rho(z,t_f)$
and its  discrete Fourier transform~\footnote{the box $L$ is chosen to be about twice the 
size of the cloud.} $\tilde{\rho}(q)$.
We acquire about 40 images  taken in the same conditions
with atom number fluctuations 
smaller than 10\%.
From this data set, we then extract the power spectrum $\langle |\tilde{\rho}(q)|^2\rangle$.
We note $\langle |\tilde{\rho}(q)|^2\rangle_i$ the power spectrum obtained before the quench
and a typical spectrum is shown  in Fig.~(\ref{fig.densityripples})$(b)$. 
We chose to normalize the momenta by $R_{\rm{TF}}^{-1}$: since the 
Fourier distribution of the $i^{\rm{th}}$ Bogoliubov mode of a 1D quasi-condensate is 
peaked at $k_i\simeq i/R_{TF}$ (see Appendix~\ref{sec.beyondLDA}), the x-axis roughly corresponds to the mode index. 
The predicted power spectrum $\langle |\tilde{\rho}(q)|^2\rangle_{\rm th}$ is computed using the LDA and analytical solution of Eq.~(\ref{eq.rhoq2}) for 
thermal equilibrium (see Appendices~\ref{SM_Density_ripples_hom}, \ref{SM_Density_ripples_LDA})%% \footnote{Note that we did not 
%% use Eq.~(\ref{eq.drasymp}) to compute 
%% \unexpanded{$\langle |\tilde{\rho}(q)|^2\rangle_{\rm th}$} since,  
%% for \unexpanded{$q R_{TF} \gtrsim 50$, $\hbar q t_v/m \gtrsim 0.5$}  and 
%% the approximation of Eq.~(\ref{eq.drasymp}) overestimates the density ripples 
%% by about 30\%.}
. This expression is peaked around 
$kR_{\rm{TF}}\simeq \sqrt{\pi m/(\hbar t_{f}})R_{\rm{TF}}\simeq 50$. 
For comparison with experimental data, we take the 
imaging resolution into account by multiplying 
$\langle |\tilde{\rho}(q)|^2\rangle_{\rm th}$ with $e^{-q^2\sigma^2}$ where $\sigma$ 
is 
the rms width of the impulse  imaging
response function, assumed to be Gaussian (Appendix \ref{sec.finite_res} discusses the effect of this finite optical resolution).
The experimental data ultimately compared well with the 
theoretical predictions, as shown in Fig.~(\ref{fig.densityripples})$(b)$, 
where   $T$
and  $\sigma$ are obtained by fitting the data~
\footnote{
The transverse size of the cloud after the time-of-flight is comparable to
the depth of focus of the imaging system and depends
on the transverse confinement. We thus expect slightly
different optical resolutions, $\sigma_i$ and $\sigma_f$ for data taken
before and after the quench respectively. We correct for
this effect to make quantitative comparison 
of data taken before and after the quench.}
\footnote{The values of resolution obtained by such fits are close to the
    expected values if one takes into account the
    depth of field of our imaging system and the fact that, after the the expansion time $t_f$
    the cloud explores  about 50$\mu$m along the imaging direction.
}.
 Finally we obtain  $k_B T/\mu_0=0.4$, close to  the
lowest value obtained in similar setups~\cite{rauer_cooling_2016,jacqmin_sub-poissonian_2011}.

\begin{figure}[h!]
\centerline{\includegraphics[width=1.05\linewidth]{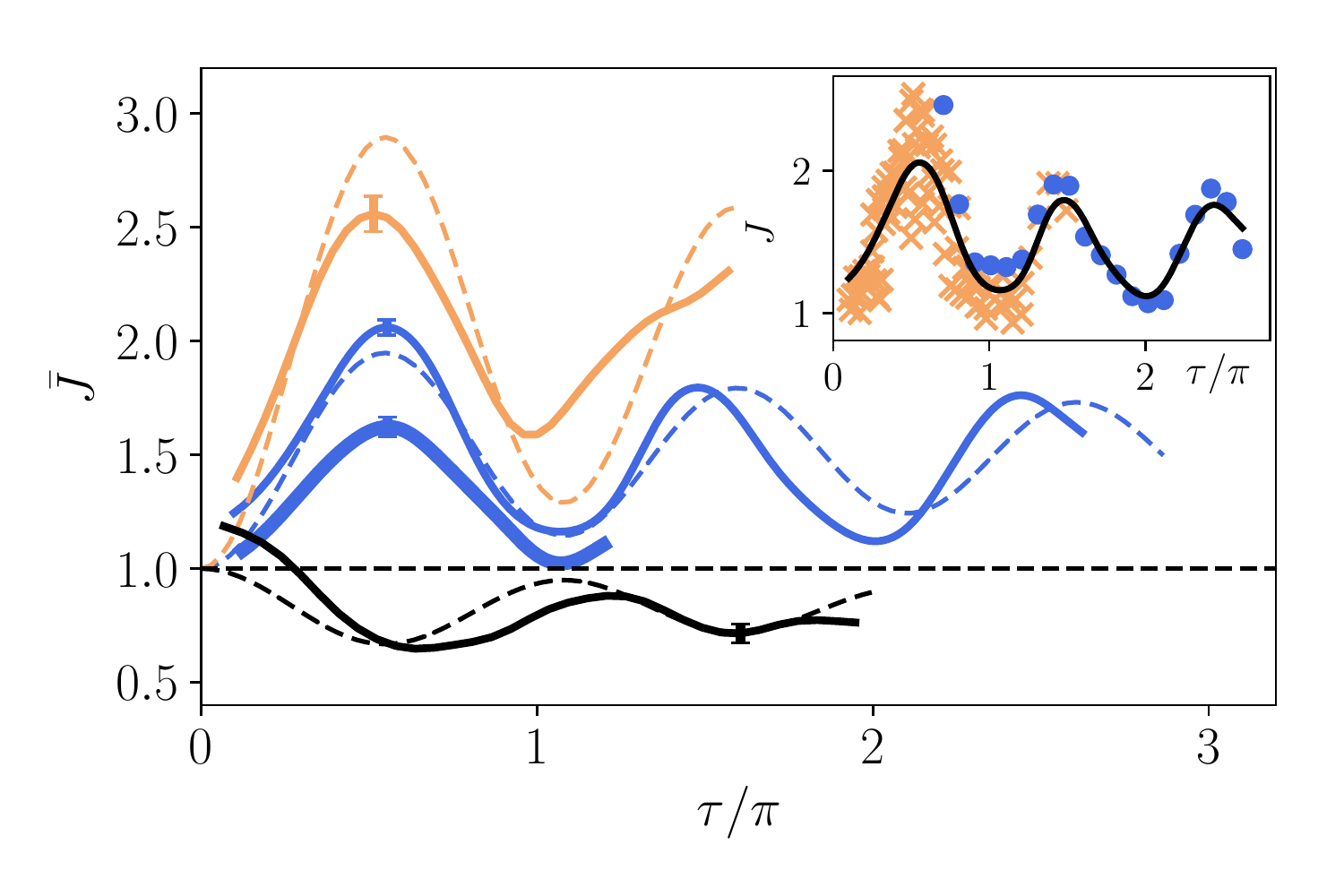}}
\caption{
Time evolution of squeezed collective modes produced by an 
interaction  quench.
The normalized  density ripples power spectrum 
is plotted versus the reduced time $\tau=cqt$, where the speed of sound $c$
is calculated for the central density. Inset shows the data corresponding to each 
measurement time and discrete $q$ values,  for a data set corresponding to 
$\kappa=2$ and $\omega_{\perp,i}=2\pi\times 1.5$~kHz, together with the resulting 
continuous averaged  quantity $\bar{J}$ 
  (see text).
Orange crosses correspond to $t<t^{g_1}_{\rm th}$ and blue circles to 
  $t>t^{g_1}_{\rm th}$. 
The main graph  shows the evolution of the experimental smoothed quantity
$\bar{J}$ for different data sets.
 The error bars show the typical statistical uncertainty on $\bar{J}$.
%The initial  transverse oscillation frequency is 1.5~kHz,
%except for the green curve for which it is 3~kHz.
The initial transverse oscillation frequency is 1.5 kHz, except for the thick dark grey (blue) curve for which it is 3~kHz.
Quench strengths are $\kappa=4$  (light gray (orange)), $\kappa=2$ (dark gray (blue)) 
data and $\kappa=-0.7$ (black). Dashed lines are theoretical 
predictions for quench strengths 
$\kappa=2$ (lightgray (orange)), 1 (light gray) and -0.35 (black).
%%   Time evolution of squeezed collective modes produced by an 
%% interaction strength.
%% The normalized  density ripples power spectrum 
%% is plotted versus the reduced time $\tau=cqt$, where the speed of sound $c$
%% is calculated for the central density. Inset shows the data corresponding to each 
%% measurement time and discrete $q$ values,  for a data set corresponding to 
%% $\kappa=2$ and $\omega_{\perp,i}=2\pi\times 1.5$~kHz, together with the resulting 
%% averaged quantity  $\bar{J}$ (see text) and the theoretical prediction
%% for $\kappa=1$ (dashed). Points in orange (crosses) correspond to $t<t^{g_1}_{\rm th}$ and blue (circles) to $t<t^{g_1}_{\rm th}$. 
%%  The main graph  shows the evolution of $\bar{J}$ for different data sets. 
%% The initial  transverse oscillation frequency is 1.5~kHz,
%% except for the fat blue curve for which it is 3~kHz.
%% Quench strengths are $\kappa=4$  (yellow data), $\kappa=2$ (fat and slim blue) 
%% data and $\kappa=-0.7$ (black data). Dashed lines are theoretical 
%% predictions for quench strengths 
%% $\kappa=2$ (orange), $1$ (blue) and $-0.35$ (black).
}
\label{fig.short}
\end{figure}

We  investigate the dynamics following the quench of the interaction strength 
by acquiring power spectra of density ripples 
at different evolution times
$t$ after the quench. We typically acquire power spectra every \unit[0.5]{ms}, over a total time
of \unit[5]{ms}. 
  A few  raw spectra are shown in Fig.~(\ref{fig.densityripples})$(b)$, 
for a quench 
strength $\kappa=2.0$. At first sight the power spectra seem erratic.
In order to reveal the expected oscillatory behavior of each Fourier component
 we introduce, for each wavevector $q$ of the discrete Fourier transform, 
and each measurement time $t$, the reduced time $\tau=cqt$, where $c$ is evaluated for the 
central density, and compute
$J(q,\tau)=\langle |\tilde{\rho}(q)|^2\rangle(t)/\langle |\tilde{\rho}(q)|^2\rangle_i$.
We restrict the range of $q$ values to $10< qR_{TF}<40$,
to ensure both  the condition $q \hbar t_f/m\ll l_c$ and the validity of 
the  LDA.
On the resulting set of spare data, shown in the inset of Fig.~(\ref{fig.short}), an oscillatory 
behavior appears, despite noise on the data.
To combine all the data in a single graph, we perform  
a ``smooth'' binning in $\tau$, {\it i.e.} we compute, for any given 
reduced time $\tau$, the weighted
averaged of the data %${J}(q,\tau)$ 
with a Gaussian weight function in $\tau$ 
of width $\Delta=0.31$ : namely
we compute 
$\bar{J}(\tau)=\sum_\alpha {J}(q_\alpha,\tau_\alpha)e^{-(\tau_\alpha-\tau)^2/(2\Delta^2)}/\sum_\alpha e^{-(\tau_\alpha-\tau)^2/(2\Delta^2)}$, 
where
the 
sum is done on the data set.
The function $\bar{J}(\tau)$, shown in Fig.~(\ref{fig.short}) 
shows a clear oscillatory behavior.

We repeat the experiment for different quench strengths 
$\kappa =(\omega_{\perp,f}/\omega_{\perp,i}-1) = \{0.3,3,5\}$, and  initial trapping 
oscillation frequencies $\omega_\perp=\{3,1.5\}~$kHz. The oscillatory behavior 
is present in all cases as shown in Fig.~(\ref{fig.short}).
We compared the observed oscillations with the theoretical predictions 
from the linearized model, Eq.~(\ref{eq.calF}). 
 The temporal behavior of the data is
in good agreement with the predicted one: both the frequency and the 
observed damping are in agreement with the predictions.
The amplitude of the experimental oscillations on the other hand are  
significantly smaller than the predictions,
and in Fig.~(\ref{fig.short}) we plot the theoretical predictions 
for quench strengths twice as small as the experimental ones.
Moreover, for a given quench strength, the observed amplitude depend on the initial 
transverse frequency, in contradiction with the theoretical model. 
Several effects leading to  a decrease of the oscillation amplitude 
are  discussed in Appendix~\ref{sec.red_ampl}. However, they  only
partially account for  the  observed amplitude reduction.

\section{Discussion}
In conclusion, analyzing density ripples, we revealed the physics at
play after a sudden quench of the interaction strength in a quasi-1D
Bose gas, namely the breathing associated to the squeezing of each
collective mode.  
The observed out-of-equilibrium dynamics 
continues for times larger than  $t^{g_1}_{\rm th}$, for 
which the $g_{1}$ function essentially reached its asymptotic 
thermal behavior~\footnote{At a time $t = t^{g_1}_{\rm th} $
    the $g_{1}$ function has reached the thermal 
value $e^{-|z|/l_c^f}$  for
    all $z < 2l_c^f$, the deviation from this thermal state 
being restricted to long distances where $g_{1}(z)<e^{-2} \approx 10\%$.}
 This can be seen in 
the inset of Fig.~(\ref{fig.short}) where data corresponding to 
$t>t^{g_1}_{\rm th}$, shown in blue circles, still present an oscillatory behavior.
This clearly underlines the 
power of the density ripple analysis to unveil out-of-equilibrium physics.
The observed damping is compatible with the sole dephasing effect due to the longitudinal harmonic confinement.
At later times, the discreteness of the spectrum and its almost constant level spacing
is expected to produce a revival phenomenon.
Its observation might however be hindered by a damping of each collective mode due to non-linear 
couplings.
 Such a damping occurs, despite the integrability of the 1D Bose gas with contact repulsive 
interactions, because the Bogoliubov collective modes do not correspond to the  conserved quantities. 
A long-lived  non-thermal nature of the state produced by the interaction strength 
might be revealed either by observing excitations in both the phononic regime and the 
particle regime of the Bogoliubov spectrum~\cite{johnson_long-lived_2017}, or, ideally, in finding a way to access the 
distribution of the Bethe-Ansatz rapidities.

\begin{acknowledgments}
This work was supported by R\'egion
\^{I}le de France (DIM NanoK, Atocirc project). 
The authors thank Dr Sophie Bouchoule of C2N (centre nanosciences 
et nanotechnologies, CNRS / UPSUD, Marcoussis, France) for the development 
and microfabrication of the atom chip. Alan Durnez and Abdelmounaim Harouri 
of C2N are acknowledged for their technical support. C2N laboratory is a 
member of RENATECH, the French national network of large facilities for 
micronanotechnology.
M. Schemmer acknowledges support by the Studienstiftung des Deutschen Volkes.
\end{acknowledgments}

%\bibliographystyle{unsrt}
%\bibliographystyle{abbrv}
%\bibliography{article-quenchg,citesm}

%%%%%%%%%%%%%%%%%%%%%%%%%%%%%%%%%%%%%%%%%%%%%%%%%%%%%%%%%%%%%%%%%%%%%%%%%%%%%%%%%%%%%%%%%%%%%%%%%%%%%%%
%%%%%%%%%%%%%%%%%%%%%%%%%%%%%%%%%%%%%%%%%%%%%%%%%%%%%%%%%%%%%%%%%%%%%%%%%%%%%%%%%%%%%%%%%%%%%%%%%%%%%%%
%%%%% Appendix%%%%%%%%%%%%%%%%%%%%%%%%%%%%%%%%%%%%%%%%%%%%%%%%%%%%%%%%%%%%%%%%%%%%%%%%%%%%%%%%%%%%%%%%
%%%%%%%%%%%%%%%%%%%%%%%%%%%%%%%%%%%%%%%%%%%%%%%%%%%%%%%%%%%%%%%%%%%%%%%%%%%%%%%%%%%%%%%%%%%%%%%%%%%%%%%

%\newenvironment{equation}{%
%\addtocounter{equation}{-1}
%\refstepcounter{defcounter}
%\renewcommand\theequation{SM\thedefcounter}
%\begin{equation}}
%{\end{equation}}

%\newpage

 \appendix

\section*{Appendix}

This appendix gives technical information and details of
calculations. 
In Appendix~\ref{sec.derivationps}  we give a general 
derivation of the density ripples power spectrum, which 
does not \textit{a priori} assume a homogeneous system.
Appendix~\ref{SM_Density_ripples_hom} gives the result for a homogeneous system and the 
analytical prediction 
for thermal equilibrium~\footnote{We corrected the formula published in~\cite{imambekov_density_2009}}.
Appendix~\ref{SM_Density_ripples_LDA} details the derivation of the density ripple power spectrum 
for a trapped gas, 
computed using the results 
for homogeneous gases and the local density approximation.
Appendix~\ref{SM_evol_LDA} provides the explicit calculation of the post-quench evolution of 
the power spectrum for a harmonically trapped gas, namely the 
calculation of the function ${\cal F}$ of the main text.
In Appendix~\ref{sec.beyondLDA} we verify the validity of the local density approximation for 
the 
parameters of the data presented in the main text. For this purpose, we compute the 
density 
ripple power spectrum using the Bogoliubov modes of the trapped gas.   
In Appendix~\ref{sec.finite_res}, we investigate the effect of finite resolution on the measured
density ripple power spectrum. We also make the link between the power spectrum and the auto-correlation
function, which permits to compare our data at thermal equilibrium with previously published work. 
In Appendix~\ref{sec.interaction_tof}, we justify that interactions play a negligible role during 
 time-of-flight, so that
the calculations of the density ripples power spectrum, which assume instantaneous 
switch-off of the 
interactions, are valid. 
In Appendix~\ref{sec.red_ampl}, we investigate two effects responsible for a reduction of the 
oscillation amplitude of the quantity $\bar{J}(\tau)$, extracted from the data, 
as compared  to the 
simple theoretical predictions Eq.~(6) of the main text: First
the finite ramp time of the interaction 
strength decreases the squeezing of the collective modes, and second 
 the finite resolution in $\tau$ 
resulting from data binning is responsible for a decrease of the
expected oscillation amplitude on the processed data.\\

\section{Derivation of the density ripples power spectrum}
\label{sec.derivationps}
The power spectrum of density ripples has been first investigated
in the limit of small density ripples and for a gas initially 
in the 3D Thomas-Fermi regime ({\it i.e.} $\mu\gg \hbar \omega_\perp$)~\cite{dettmer_observation_2001,hellweg_phase_2001}. 
It was then computed assuming instantaneous switching off of the interactions 
in~\cite{imambekov_density_2009}. Here, for consistency, we 
rederive Eq.~(4) and (5) of the main text.
Since we will later consider trapped gases, let us first assume a general scenario where we do not restrict ourselves to the homogeneous case.
We let the gas evolve freely for a time $t_f$
after interactions have been switched off. 
The power spectrum of the density fluctuations  after $t_f$ writes 
\begin{equation}
\langle |\tilde{\rho}(q)|^2\rangle =\int\int dz_1 dz_2 e^{iq(z_1-z_2)}\langle \delta n(z_1,t_f)\delta n(z_2,t_f)\rangle.
\label{eq.powersectrumdef}
\end{equation}
Writing $\delta n(z)=n(z)-\langle n(z)\rangle$ and expanding the above equation, the
term $|\int dz e^{iqz}\langle n(z,t_f)\rangle|^2$ appears. 
Here we consider times of flight
short enough so that the shape of the cloud barely changes during 
time of flight, so that $\langle  n(z,t_f)\rangle\simeq \langle  n(z,0)\rangle$. We moreover 
consider wavevectors $q$ much larger than the inverse of the cloud length, such that 
$|\int dz e^{iqz} \langle  n(z,0)\rangle|^2 $  is a negligible quantity. We then have
 
\begin{equation}
\langle |\tilde{\rho}(q)|^2\rangle  \simeq \int\int dz_1 dz_2 e^{iq(z_1-z_2)}\langle n(z_1,t_f) n(z_2,t_f)\rangle.
\label{eq.pstf}
\end{equation}
To compute $ n(z,t_f) = \Psi^{+}(z,t_f)\Psi(z,t_f)$ we evolve the atomic field with the free-particle propagator, which leads to
\begin{equation}
\psi(z,t_f)=\frac{1}{\sqrt{2\pi t_f}}\int d\alpha \psi(\alpha,0) e^{i\frac{(z-\alpha)^2}{2t_f}}, 
\end{equation}

where for simplicity we use a unit system in which $m=\hbar = 1$.
We then have
\begin{widetext}
\begin{equation}
\langle n(z_1,t_f) n(z_2,t_f)\rangle=
\frac{1}{(2\pi t_f)^2}\int\int\int\int d\alpha d\beta d\gamma d\delta 
\langle \psi^+_\alpha\psi_\beta \psi^+_\gamma \psi_\delta \rangle 
e^{-i\frac{(z_1-\alpha)^2}{2t_f}}
e^{i\frac{(z_1-\beta)^2}{2t_f}}
e^{-i\frac{(z_2-{\gamma})^2}{2t_f}}
e^{i\frac{(z_2-\delta)^2}{2t_f}},
\end{equation}

where we use the simplified notation $\psi_\nu=\psi(\nu,0)$. 
Expanding the exponentials, the above expression writes

\begin{equation}
\langle n(z_1,t_f) n(z_2,t_f)\rangle=
\frac{1}{(2\pi t_f)^2}\int\int\int\int d\alpha d\beta d\gamma d\delta 
\langle \psi^+_\alpha\psi_\beta \psi^+_\gamma \psi_\delta \rangle 
e^{i\frac{(\alpha-\beta)z_1}{t_f}}
e^{i\frac{\beta^2-\alpha^2}{2t_f}}
e^{i\frac{(\gamma-\delta)z_2}{t_f}}
e^{i\frac{\delta^2-\gamma^2}{2t_f}}.
\end{equation}

Injecting into Eq.~(\ref{eq.pstf}), and using $\int dz e^{ikz}=2\pi\delta(k)$ and 
$\delta(x/\alpha)=\alpha\delta(x)$, we get
\begin{equation}
\langle |\tilde{\rho}(q)|^2\rangle=
\int\int d\alpha  d\delta 
\langle \psi^+_\alpha\psi_{\alpha+qt_f} \psi^+_{\delta + qt_f} \psi_{\delta} \rangle 
e^{-i\frac{\alpha^2}{2t_f}}
e^{i\frac{(\alpha+qt)^2}{2t_f}}
e^{-i\frac{(\delta+qt)^2}{2t_f}}
e^{i\frac{\delta^2}{2t_f}}.
\end{equation}

Defining $\delta = \alpha+X$, we obtain

\begin{equation}
\langle |\tilde{\rho}(q)|^2\rangle=
\int\int d\alpha  dX e^{iqX} 
\langle \psi^+_\alpha\psi_{\alpha+qt_f} \psi^+_{\alpha+X + qt_f} \psi_{\alpha + X} \rangle .
\end{equation}
For gases lying deep in the quasi-condensate regime, one can neglect density fluctuations 
when estimating the expectation value in the above equation, such that 
\begin{equation}
\langle |\tilde{\rho}(q)|^2\rangle
 \simeq\int\int d\alpha  dX e^{iqX} 
\sqrt{n(\alpha)n(\alpha+qt_f)n(\alpha+X + qt_f)n(\alpha + X)}
\langle e^{i(\theta(\alpha) - \theta({\alpha+qt_f}) + \theta({\alpha+X + qt_f}) - \theta({\alpha + X}))}\rangle.
\label{eq.powerspectrumgeneral}
\end{equation}
\end{widetext}

The following section applies this result to homogeneous systems. 
This equation is however 
not restricted to homogeneous systems and we will use it to treat the
effect of the trap beyond the local density approximation.

\section{Power spectrum of the density ripples for a 
homogeneous gas}\label{SM_Density_ripples_hom}
For a homogeneous gas, the relevant quantity is 
an intensive variable which  relates to the 
expression $\langle |\tilde{\rho}(q)|^2\rangle$ of the previous 
section  by
\begin{equation}
\langle |\rho(q)|^2\rangle=\frac{1}{L}\langle |\tilde{\rho}(q)|^2\rangle
\label{eq.rhotildevsrho}
\end{equation} 
where   $L$  is the length of the box.  
Injecting   Eq.~(\ref{eq.powerspectrumgeneral})
into Eq.~(\ref{eq.rhotildevsrho}), we recover
Eq.~(3) and (4) of the main text, up to an irrelevant term in $\delta(q)$~\footnote{This term is due to the approximation made when going from 
Eq.~(\ref{eq.powersectrumdef}) to 
Eq.~(\ref{eq.pstf}), which is valid only for $q$ values larger than the inverse of the 
cloud size.}.
In fact, Wick's theorem is applicable since $\theta$ is a Gaussian variable
\footnote{Since the 
Hamiltonian of interest is quadratic in $\theta$, the distribution of $\theta$ is 
Gaussian at thermal equilibrium. The squeezing of each collective mode
produced by the interaction quench 
preserves the Gaussian nature of $\theta$.}, which leads to 
\begin{equation}
\frac{\langle |{\rho}(q)|^2\rangle}{n_0^2}=
  \int   dX e^{iqX -\frac{1}{2}\langle (\theta(0) - \theta({qt_f}) + \theta({X + qt_f}) - \theta({ X}))^2\rangle}.
\label{eq.powerspectrumhomogeneous}
\end{equation}

\begin{figure}
  \centerline{\includegraphics[height=4cm]{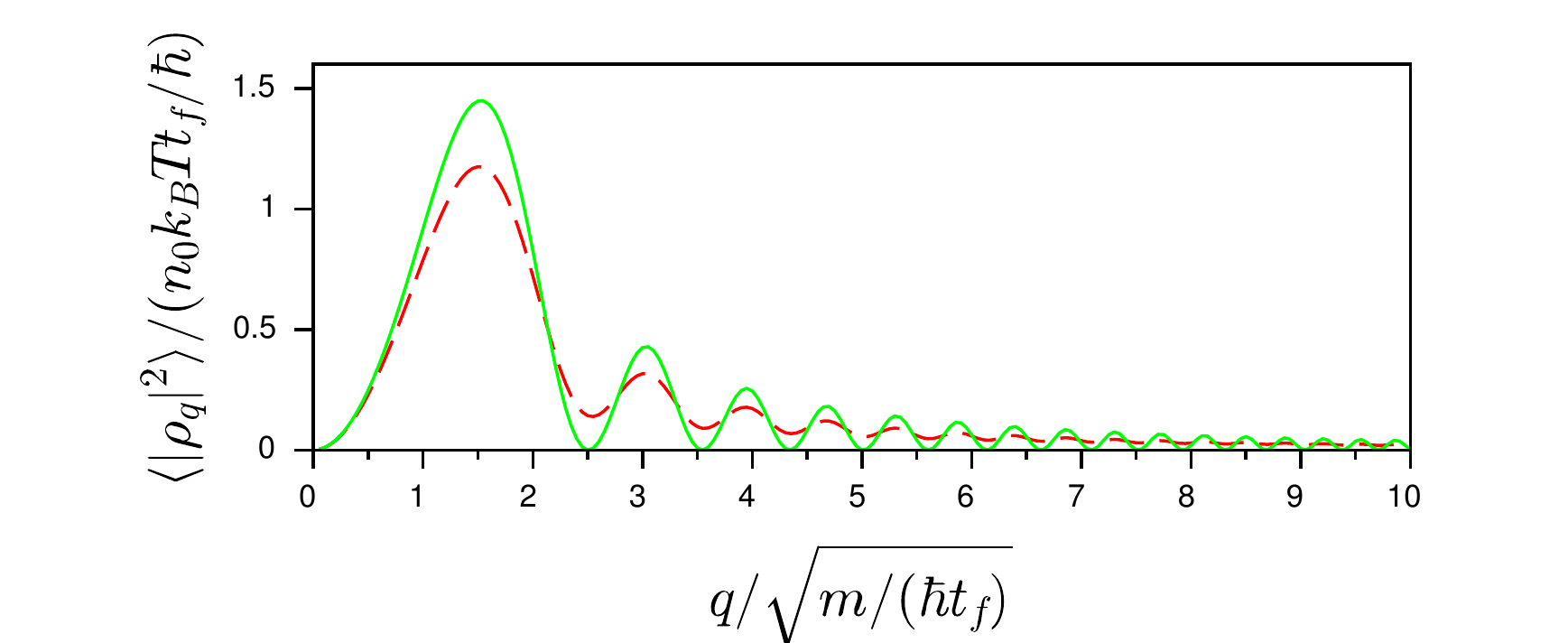}}
  \vspace{0.2cm}
  
\caption{Density ripples power spectrum for a homogeneous gas.
The exact formula Eq.~(\ref{eq.rhoq2thermohom}) (dashed curve) is compared to the small $q$ approximation
given Eq.~(5) of the main text, 
where $\langle |{\rho}(q)|^2\rangle$ is proportional to $\langle \theta_q^2\rangle$ (solid curve).
The only relevant parameter is $\hbar t_{f}/(ml_c^2)$. Results 
are shown for $\hbar t_{f}/(ml_c^2)=0.05$, a value corresponding to 
 the data depicted in Fig. (2,b) of the main text, the correlation length 
 $l_c=2\hbar^2n_0/(mk_BT)$ being computed for  the central density.
   The effect of the imaging resolution
is to multiply this theoretical power spectrum 
with $e^{-\sigma^2 q^2}$, where $\sigma$ is the rms width of the 
imaging pulse response function, assumed to be Gaussian.
For our data, $\sigma\sqrt{m/(\hbar t_f)}=0.85$ and only the first maximum
of $\langle |{\rho}(q)|^2\rangle$ remains visible.
}
\label{fig.testqpetit}
\end{figure}

To compute the power spectrum of density ripples for a thermal equilibrium 
state, we follow the calculation made in~\cite{imambekov_density_2009} 
and expand
the exponential term in Eq.~(\ref{eq.powerspectrumhomogeneous}) 
as a function of the
first order correlation
function $g^{(1)}(z)= n_0e^{-\frac{1}{2}\langle (\theta(0) - \theta(z))^2\rangle }$,
which fulfils $g^{(1)}(z)=n_0e^{-|z|/l_c}$ where $l_c= 2 \hbar^2 n_0/(k_B T)$~\cite{imambekov_density_2009}.
Calculation of the integral in Eq.~(\ref{eq.powerspectrumhomogeneous}) then leads to
%\begin{widetext}
\begin{multline}
\frac{\langle |{\rho}(q)|^2\rangle}{n_0^2} =
  \frac{4 ql_c}{q(4+l_c^2q^2)} \\
  - \frac{ 4 e^{- \frac{2\hbar qt_f}{ml_c}}\left ( q l_c \cos(\frac{\hbar q^2t_f}{m})+2 \sin(\frac{\hbar q^2t_f}{m})\right )}{q(4+l_c^2q^2)}.
\label{eq.rhoq2thermohom}
\end{multline}
%\end{widetext}
Note that we corrected the formula given in~\cite{imambekov_density_2009}.
The power spectrum computed with this equation is compared in Fig.~\ref{fig.testqpetit}
to the approximated formula valid for small $q$, namely Eq.~(5) of the main text. 

\section{Density ripple power spectrum for a harmonically confined gas under the LDA}\label{SM_Density_ripples_LDA}
Let us investigate the density ripples power spectrum in the case of a gas trapped in a longitudinal
potential smooth enough so that the cloud size $L$ is much larger than the typical phase
correlation length $l_c$ and much larger than $\hbar q t_f/m$: $L\gg l_c,\hbar q t_f/m$. 
As in section~\ref{sec.derivationps}, we moreover consider the 
power spectrum for wavevectors $q\gg 1/L$.
 Let us start with the general expression Eq.~(\ref{eq.powersectrumdef}) that we write
 \begin{equation}
 \langle |\tilde{\rho}(q)|^2\rangle =\int dz \int du 
\langle \delta \rho(z,t_f)\delta \rho(z+u,t_f)\rangle e^{iqu}.
\end{equation}
%Since we are interested in wavevectors $q\gg 1/L$, and 
Consider $\langle \delta \rho(z,t_f)\delta \rho(z+u,t_f)\rangle$ 
for a given $z$. This expression vanishes 
over a length much smaller than $L$, so values of $u$ significantly contributing to the integral are much smaller than $L$.
Moreover the region of the initial cloud contributing most to  $\langle \delta \rho(z,t_f)\delta \rho(z+u,t_f)\rangle$
is much smaller than $L$ for sufficiently large $L$. 
Then, to compute  $\langle \delta \rho(z,t_f)\delta \rho(z+u,t_f)\rangle$
one can perform a local density approximation and use the 
result of a homogeneous 
gas at a density
$n_0(z)$.
We then obtain
\begin{equation}
\langle |\tilde{\rho}(q)|^2\rangle=\int dz \langle |{\rho}_{n_0(z)}(q)|^2\rangle
\label{eq.LDAint}
\end{equation}
where the subscript $n_0(z)$  specifies that one considers the result for a homogeneous 
gas of density $n_0(z)$. This expression is referred to as the local density approximation
expression (LDA) of the power spectrum.
We have tested this approximation, for conditions close to
the experimental data
presented in the main text, by comparing it with calculations based on the Bogoliubov excitations
of the trapped system (see section \ref{sec.beyondLDA}).

\section{Time evolution of the density ripple power spectrum for a harmonically confined gas}\label{SM_evol_LDA}
\begin{figure}
%\centerline{\input{Fronde.tex}}
\includegraphics[width = \linewidth]{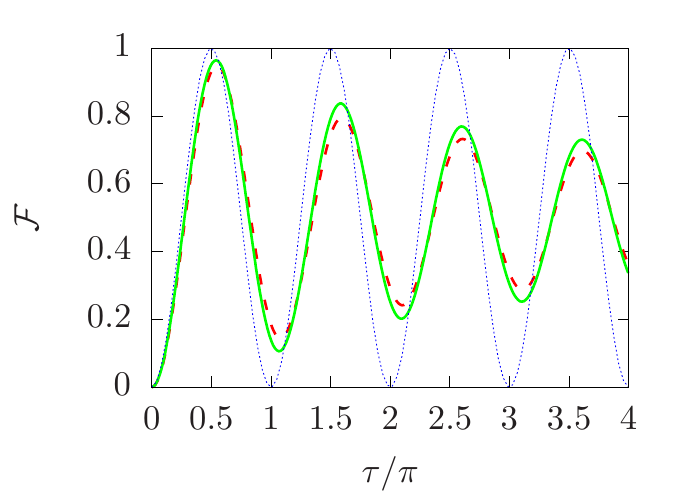}
\caption{Oscillation of each spectral component of the power spectrum for a harmonically confined 
gas in the LDA (color online). The function  ${\cal F}$ is shown in thick solid  lines (green), for $\eta=\mu_p/(\hbar \omega_\perp)=1.0$.
The pure 1D limit, corresponding to $\eta\ll 1$ is shown as dashed (red) lines. The 
undamped oscillations expected for a homogeneous gas are shown in dotted (blue) line. In all the cases, $\tau=cqt$ where 
$c$ is the central sound velocity.}
\label{fig.Fronde}
\end{figure}

Here we give an explicit derivation of Eq.~(6) of the main text, for a gas harmonically 
confined in a longitudinal trap of frequency $\omega_\parallel$.
Injecting Eq.~(5) and Eq.~(2) of the main text into 
Eq.~(\ref{eq.LDAint}), 
and using the local initial
power spectrum of $\theta$ which writes 
$\langle \theta_q^2\rangle =mk_B T /(\hbar^2 n_0 q^2)$, we derive
Eq.~(6) of the main text with 
\begin{equation}
{\cal F}= \int dz n_0(z) \sin^2 \left ( c(z) q t \right )/N
\label{eq.Frondeint}
\end{equation}
where $N$ is the total atom number. 
The density profile $n_0(z)$ is estimated 
itself within the LDA, using the local chemical 
potential 
\begin{equation}
\mu(z)=\mu_p(1-(z/R_{\rm{TF}})^2)),
\label{eq.muloc}
\end{equation}
where $R_{\rm{TF}}$ is the Thomas-Fermi radius of the density profile and $\mu_p$ is the chemical 
potential at the trap center.
%% The local speed of sound fulfills
%% \begin{equation}
%% c=\sqrt{\frac{n\partial \mu /\partial n}{m}}.
%% \end{equation}
For a transverse harmonic confinement  of frequency $\omega_\perp$,
it has been checked, 
by comparing with predictions of the 3D Gross-Pitaevskii equation,
that the equation of state of the gas 
is very well described by the heuristic formula~\cite{fuchs_hydrodynamic_2003}
\begin{equation}
\mu(n)=\hbar\omega_\perp\left ( \sqrt{1+4na} -1\right),
\label{eq.eos}
\end{equation}
where $a$ is the 3D scattering length between atoms.
For small linear densities, we recover the 1D expression $\mu=2\hbar \omega_\perp a n$, 
valid far from the confinement-induced resonance~\cite{olshanii_atomic_1998}. 
Using Eq.~(\ref{eq.eos}) and Eq.~(\ref{eq.muloc}), we obtain the density profile
\begin{equation}
n_0(z)=\left [ \left ( \eta (1-\tilde{z}^2)+1\right )^2 -1 \right ]/(4a)
\end{equation}
where we introduced $\tilde{z}=z/R_{\rm{TF}}$ and $\eta=\mu_p/(\hbar\omega_\perp)$.
This yields $N=(4/3\eta+8\eta^2/15)R_{\rm{TF}}/(2a)$.
The local speed of sound on the other hand, obtained from the
thermodynamic relation $c=\sqrt{n(\partial \mu /\partial n)/m}$,  writes
\begin{equation}
c(z)=c_p \sqrt{\frac{(1+\eta)\left [  \left ( 1+\eta(1-\tilde{z}^2)\right)^2-1\right]}
{\left (1+\eta(1-\tilde{z}^2) \right) \left ( (1+\eta)^2-1\right)}},
\end{equation}
where $c_p$ is the speed of sound computed for the central density.
Injecting into Eq.~(\ref{eq.Frondeint}), we then find
\begin{widetext}
\begin{equation}
{\cal F}=\frac{1}{4\eta/3+8\eta^2/15}\int_0^1 d{\tilde z} 
\left [ \left ( 1+\eta(1-\tilde{z} ^2)\right)^2 -1 \right ]
\sin^2\left ( \tau \sqrt{\frac{(1+\eta)\left [  \left ( 1+\eta(1-\tilde{z}^2)\right)^2-1\right]}
{\left (1+\eta(1-\tilde{z}^2) \right) \left ( (1+\eta)^2-1\right)}}  \right ).
\end{equation}
\end{widetext}
When the gas is deeply 1D, namely for $\eta\ll1$, this expression reduces to 
\begin{equation}
{\cal F}_{\rm{1D}}=\frac{3}{2}\int_0^1 d{\tilde z} 
(1-\tilde{z} ^2)
\sin^2\left ( \tau\sqrt{1-\tilde{z}^2}\right ).
\end{equation}
Experimentally, values of $\eta$ are in the range $[0.6; 1.3]$. 
Fig.~\ref{fig.Fronde} shows the function ${\cal F}$, computed for $\eta=1$. 
 We compare it to  ${\cal F}_{\rm{1D}}$  and to the expression expected 
for a homogeneous gas, namely
$\sin^2(\tau)$.

\section{Beyond the LDA: calculation using 
Bogoliubov modes of a harmonically confined 1D gas}
\label{sec.beyondLDA}
%\subsection{Phononic Bogoliubov modes for hamronically trapped AD gas}
Here we consider a 1D gas confined longitudinally in a harmonic 
trap of frequency $\omega_\parallel$. In opposition to the calculations done in the 
previous section we do not rely on the local density approximation but use the 
Bogoliubov modes of the trapped gas to compute the post-quench evolution and the 
density ripples power spectrum.
The relevant collective modes lie deep in the phononic regime. 
The Bogoliubov modes, indexed by an integer $\nu$, 
then acquire an analytical dispersion relation and 
analytical wavefunctions that one can use for calculations.
For each mode, the dynamics are accounted for by the harmonic oscillator Hamiltonian 
\begin{equation}
H_\nu=\hbar\omega_\nu\left (\frac{ x_{\nu}^2}{2} +\frac{p_{\nu}^2}{2}\right ),
\label{eq.Hnu}
\end{equation}
where $\omega_\nu =\omega_\parallel \sqrt{\nu(\nu+1)/2}$
and $x_{\nu}$ and $p_{\nu}$ are canonically conjugate variables. 
The phase and density fluctuation operators write
\begin{equation}
\left \{ 
\begin{array}{l}
\theta(z)=\sum_\nu {\theta_\nu}(z) p_\nu \\
\delta n(z)=\sum_\nu {n_\nu}(z) x_\nu \\
\end{array}
\right .
\label{eq.thetavspnu}
\end{equation}
where
\begin{equation}
\left \{ 
\begin{array}{l}
{\theta_\nu}(z)=\frac{1}{\sqrt{2}}\left (\frac{mg}{\hbar^2n_p}\right )^{1/4}\frac{\sqrt{2\nu+1}}{\left ( \nu(\nu+1) \right )^{1/4}}
P_\nu(\frac{z}{R_{TF}})    \\
{n_\nu}(z)=\frac{\sqrt{2\nu+1}}{2R_{\rm{TF}}}\left (\nu(\nu+1) \right )^{1/4} \left( 
\frac{\hbar^2n_p}{mg}\right )^{1/4} 
P_\nu(\frac{z}{R_{TF}}).
\\
\end{array}
\right .
\label{eq.thetavsLegendre}
\end{equation}
Here $n_p$ and $R_{TF}$ are the central density and radius of the Thomas-Fermi profile 
$n_0(z)=n_p(1-(z/R_{TF})^2)$ and $P_\nu$ are the Legendre polynomials.
The interaction quench consists 
of a sudden change of the interaction parameter 
$g$ from 
$g_i$ to $g_f=(1+\kappa)g_i$ at $t=0$, while changing the longitudinal oscillation
frequency by a factor $\sqrt{1+\kappa}$ so that $R_{\rm{TF}}$ stays constant.  
Then the interaction quench preserves 
the shapes of the wavefunctions $\theta_\nu$ and $n_\nu$, and it simply 
changes the canonical variables $x_{\nu}$ and $p_{\nu}$ according to 
\begin{equation}
\left \{
\begin{array}{l}
x_{\nu}(t=0^+)=(g_f/g_i )^{1/4}x_{\nu}(t=0^-)\\
p_{\nu}(t=0^+)=(g_i/g_f )^{1/4}p_{\nu}(t=0^-)
\end{array}
\right .
\label{eq.squeezingBogo}
\end{equation}
Under such a transformation, the initial thermal state, an isotropic Gaussian, 
becomes a squeezed state and
its subsequent evolution under the Hamiltonian Eq.~(\ref{eq.Hnu}) 
leads to a breathing of each quadrature.
In particular 
\begin{equation}
\langle p_\nu^2\rangle=\langle p_\nu^2\rangle_i
\left (
1+\kappa\sin^2(\omega_\nu t)
\right ).
\label{eq.evolthetanu}
\end{equation}
The initial value $\langle p_\nu^2\rangle_i$ is given by the thermal expectation value, which reduces to 
\begin{equation}
\langle p_\nu^2\rangle_i=k_B T /(\hbar \omega_\nu)
\label{eq.pnueq}
\end{equation}
for the low-lying modes for which $k_B T \gg \hbar \omega_\nu$.

Injecting Eq.~(\ref{eq.thetavspnu}) into Eq.~(\ref{eq.powerspectrumgeneral}),
using Wick's theorem and 
the fact that different modes are uncorrelated we get
\begin{widetext}
\begin{equation}
\langle |\tilde{\rho}(q)|^2\rangle = 
\int\int d\alpha dX \begin{array}[t]{l}
e^{iqX}\sqrt{n_0(\alpha)n_0(\alpha+qt_f)n_0(\alpha+X + qt_f)n_0(\alpha + X)}\\
e^{-\frac{1}{2}
\sum_\nu \langle p_\nu^2\rangle \left ( \theta_\nu(\alpha) - \theta_\nu({\alpha+qt_f}) + 
\theta_\nu({\alpha+X + qt_f}) - \theta_\nu({\alpha + X})\right )^2}.
\end{array}
\label{eq.psLegendre}
\end{equation}
\end{widetext}
For $\hbar q t_f/m\ll l_c$, where $l_c$ is the phase correlation length, 
one can expand the exponential and $\langle |\tilde{\rho}(q)|^2\rangle$
is obtained by summing the contribution of each mode. Since the Legendre
polynomials behave as $\cos((\nu+1/2)x+\pi/4)$ at small $x$, 
the contribution of 
the mode $\nu$ is peaked at $q\simeq \nu/R_{TF}$. 

\begin{figure}
  \centerline{%\begin{tabular}{p{8cm}p{8cm}}
    %\centerline{
    \includegraphics{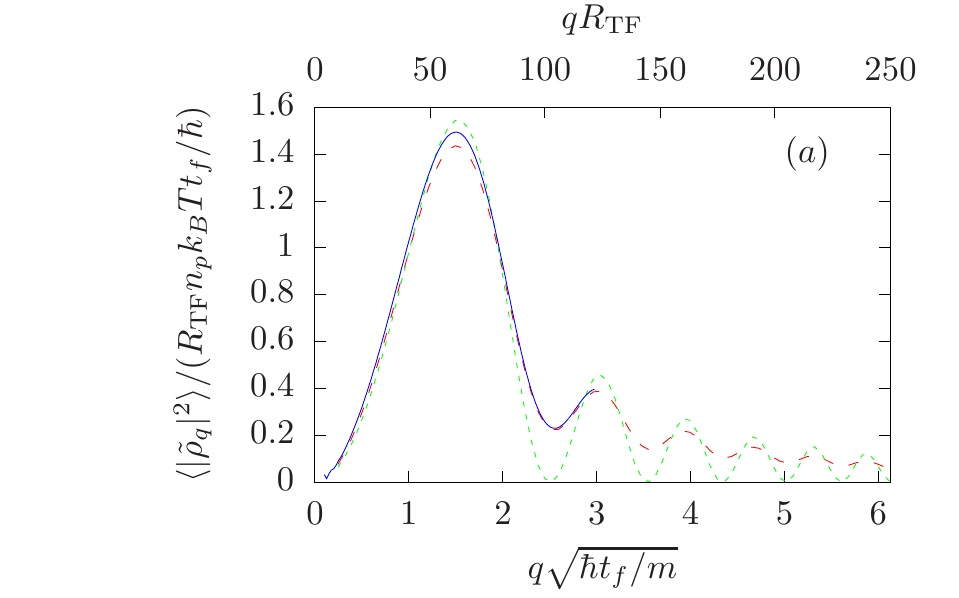}%}&
%      \centerline{\includegraphics{faitfigurethermotheores.eps}}\\
 %     \end{tabular}
  }
\caption{Test of the local density approximation (LDA) (color online).
The plot shows the  density ripples spectrum of a gas at thermal 
equilibrium confined in a harmonic potential. 
The complete calculation, based on the expansion on the Bogoliubov modes, 
whose wavefunctions are given by the Legendre polynomial, is shown in 
solid line (blue).
It is in excellent agreement with the spectrum computed within the 
local density approximation (LDA) shown in dashed line (red).
The further approximation of small wavevectors, Eq.~(5) of the main text, 
injected into
the LDA,  shown in dotted line (green),
is also in good agreement, for wavevectors fulfilling
$qR_{\rm{TF}}<50$.
Calculations are done for a Thomas-Fermi radius  $l_c/R_{\rm{TF}}=0.2$ and 
and time-of-flight $t_f= 0.015 ml_c^2/\hbar$,
where $l_c=2\hbar^2n_p/(mk_BT)$ is the correlation length 
at the center of the cloud. These parameters are close to those of the
experimental data.
}
\label{fig.eqthermo}
\end{figure}

The predictions of Eq.~(\ref{eq.psLegendre}) may 
be compared to the one obtained within the 
Local density approximation. 
Here we focus on the case of thermal equilibrium.
We compute the density ripple spectrum injecting  
the thermal equilibrium value Eq.~(\ref{eq.pnueq})
and the mode wavefunction Eq.~(\ref{eq.thetavsLegendre})
into 
Eq.~(\ref{eq.psLegendre}).
Fig.~\ref{fig.eqthermo} shows the result for 
a cloud whose Thomas-Fermi radius fulfils $l_c/R_{\rm{TF}}=0.2$, 
where $l_c=2 \hbar^2 n_p/(mk_BT)$ is the correlation length 
of the first order correlation function at the center of the cloud, 
and for a time-of-flight $t_f= 6\times 10^{-4} mR_{\rm{TF}}^2/\hbar$.
These parameters are close to the experimental ones.
We compared the results with the LDA, together with the 
analytical formula for homogeneous gases Eq.~(\ref{eq.rhoq2thermohom}) and
we find excellent agreement.
We also compare with the LDA but using, 
instead of Eq.~(\ref{eq.rhoq2thermohom}), the approximation Eq.~5 of the main text.
We find very good agreement as long as $qR_{\rm{TF}}<50$.

\section{Effect of a finite optical resolution and auto-correlation function}\label{sec.finite_res}
\begin{figure}
      \centerline{\includegraphics{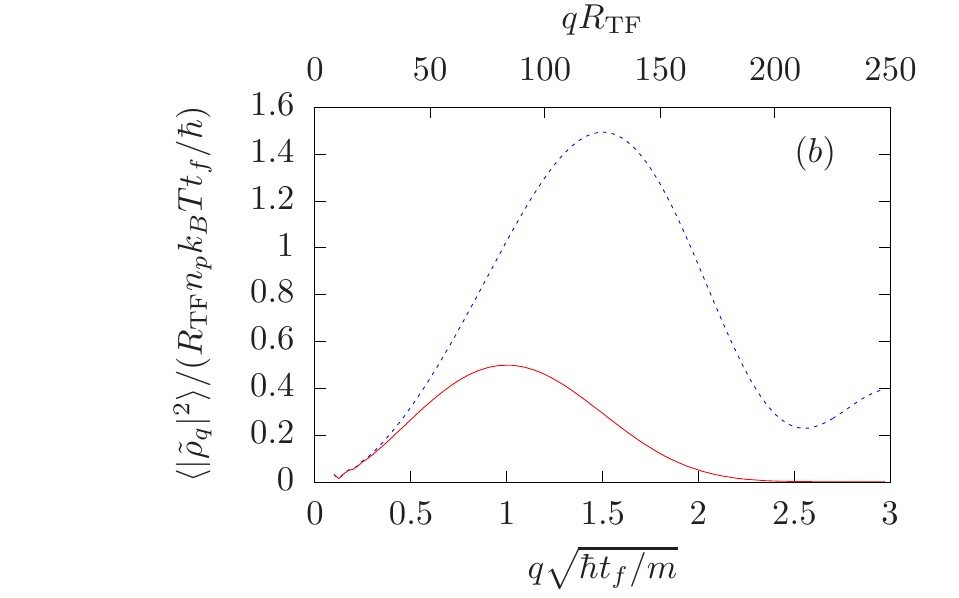}}
\caption{Effect of the finite resolution (color online). 
  We consider a clould at thermal equilibrium in a harmonic
  potential with the same parameters as in Fig.~\ref{fig.eqthermo}.
  The power spectrum for infinite resolution (blue dashed curve) is
  compared to the power spectrum expected for a finite imaging resolution (red solid curve).
  The effect of the imaging resolution
is to multiply the power spectrum 
with $e^{-\sigma^2 q^2}$, where $\sigma$ is the rms width of the 
imaging pulse response function, assumed to be Gaussian. Here we took
$\sigma\sqrt{m/(\hbar t_f)}=0.85$, a value close to that of
experimental data.}
\label{fig.resolution}
\end{figure}

 The effect of the imaging resolution is to multiply the theoretical power spectrum of density ripples
with $e^{-\sigma^2 q^2}$, where $\sigma$ is the rms width of the 
imaging pulse response function, assumed to be Gaussian.
The resulting power spectrum, for a harmonically confined cloud at thermal equilibrium,
is shown in Fig.~(\ref{fig.resolution}) for
$\sigma\sqrt{m/(\hbar t_f)}=0.85$, a value
typical for our experiments.
 The large $q$ behavior of the power spectrum is highly dominated by the effect of resolution
and only the first maximum of $\langle |{\rho}(q)|^2\rangle$ remains visible.
Fitting the experimental power spectrums for clouds at thermal equilibrium,
we extract both the temperature and the imaging resolution (see Fig.~(2) of the main text).
 The obtained rms widths $\sigma$, close to 3~$\mu$m, are compatible with the 
expected values if one takes into account the
depth of focus of our imaging system ($\simeq 5\mu$m)
and the fact that, after the the expansion time $t_f$
the cloud explores several tens of $\mu$m along the imaging axis.
Note finally that the  imaging resolution is irrelevant for the
investigation of the dynamics following an interaction quench, since, for each Fourier
component $q$, we investigate the time behavior of  the normalised quantity
$\langle |\tilde{\rho}(q)|^2\rangle(t)/\langle |\tilde{\rho}(q)|^2\rangle_i$ (see main text):
the imaging resolution has no effect on this normalised quantity. 

\begin{figure}
\includegraphics[width = \linewidth]{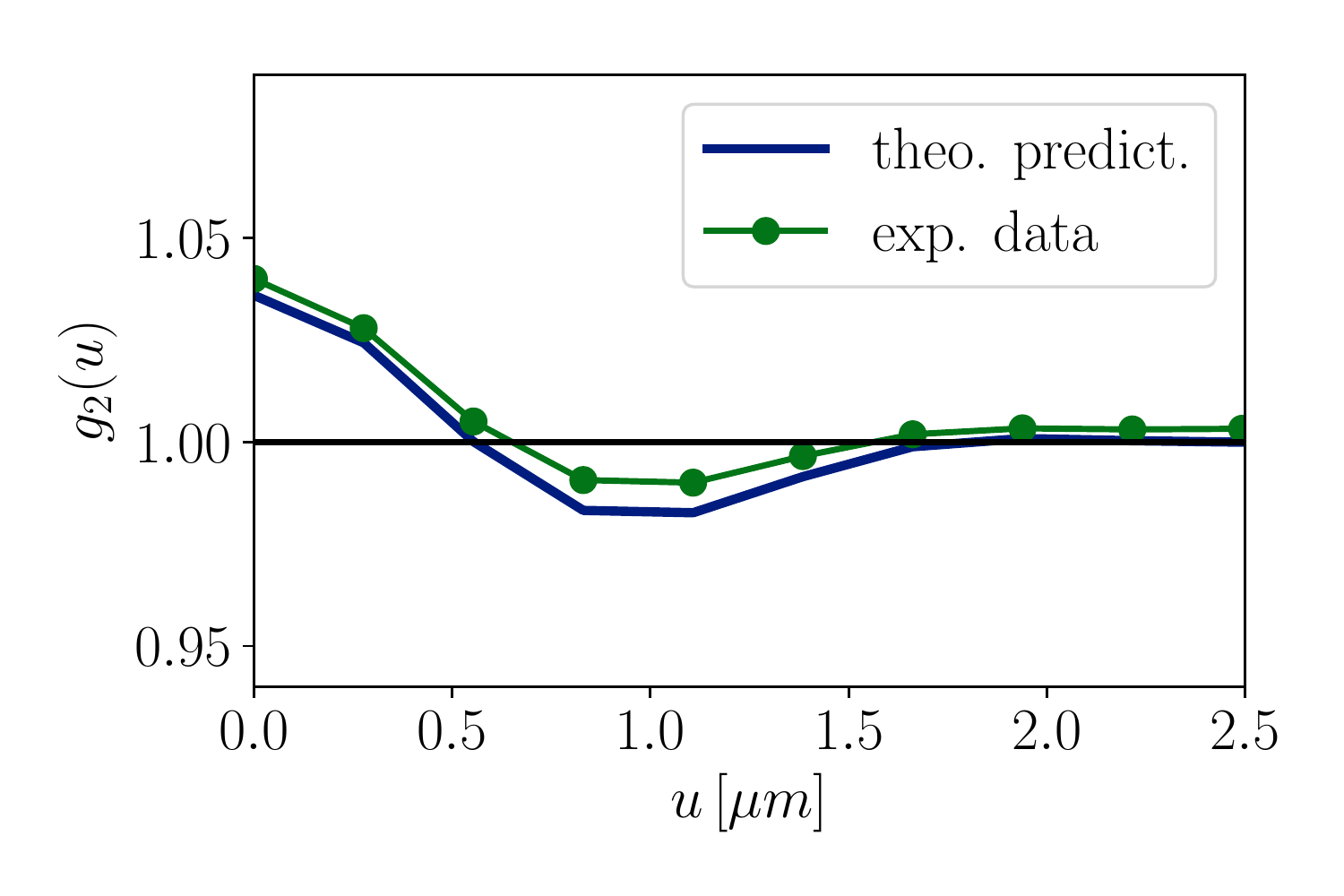}
      %\centerline{\includegraphics{faitfigurethermotheores.eps}}
  \caption{Normalised auto-correlation function of the density ripples.
    The data set used is the same as that
  of Fig. (2)(b) of the main text. Experimental data are shown in green
  and the theoretical prediction for a cloud at a temperature $T=55$ nK and an optical resolution
  $\sigma=2.9~\mu$m is shown in blue.}
\label{fig.auto_correlation}
\end{figure}

In our paper, we extract from the data the density ripple power spectrum since it is the
relevant quantity that enable to resolve the collective Bogoliubov modes.
Alternatively, one could consider the auto-correlation function of the density ripples
$C(u)=\int dz \langle \delta n(z)\delta n(z+u)\rangle dz$, which
is the Fourier transform of the density ripple power spectrum:
$C(u)=1/(2\pi)\int dq \langle |\tilde{\rho}(q)|^2\rangle e^{-iqu}$.
In~\cite{manz_two-point_2010}, the authors introduced the normalised auto-correlation function
$g_2(u)=1+C(u)/\int du\langle n(z)\rangle\langle n(z+u)\rangle$.
Fig.~(\ref{fig.auto_correlation}) shows $g_2(u)$  for the data at thermal equilibrium (before the quench)
shown in Fig.~(2) of the main text. A behavior very similar to that observed in~\cite{manz_two-point_2010} is recovered. 

\section{Beyond instantaneous interaction switch off: finite transverse expansion time}\label{sec.interaction_tof}
In the data presented in the main text, the frequency of the probed longitudinal modes, of the order
of $cq$, is no more than $0.15\times\omega_\perp$. Then,
due to the rapid transverse expansion,  interactions
during time-of-flight become almost instantaneously
negligible and are expected to give only minor corrections to the
density ripples spectrum computed for an instantaneous switching off of 
the interactions. 
It is nevertheless interesting to estimate their effect. 
This has already been computed 
in~\cite{hellweg_phase_2001}, in the  limit $\mu\gg\hbar\omega_\perp$ and 
using time-dependent Bogoliubov equations, {\it i.e.} equations of motion 
linearized in 
density fluctuations and phase gradient.
The linearized calculations {\it a priori} require that density fluctuations stay small.
Although in our case density ripples at the end of the time-of-flight have large 
amplitudes, the
Bogoliubov calculations hold for the small $q$ components, which fulfil $ q\ll m l_c/(\hbar   t_f)$  
and  which are considered in our paper. 
The condition $\mu\gg\hbar\omega_\perp$ on the other hand  
is not verified for the data shown in 
the main text. 
 We nevertheless 
believe that the calculations of~\cite{hellweg_phase_2001} 
give a relevant estimation of the 
effect of interactions during the time-of-flight for our data.
 From results of~\cite{hellweg_phase_2001}, 
we find that the density ripples power spectrum for the small $q$ 
wavevectors, given by equation~(5) of the main text,  should be corrected by the factor 
\begin{equation}
\mathcal{C} = \left( \omega_{\perp} t_f\right)^{-{\left( \frac{c q}{\omega_{\perp}}\right)^2}}.
\end{equation}
In all experimental situations $ \mathcal{C} > 0.95 $, which confirm that 
the effect of interactions during the time-of-flight is small. 

\section{Effects which may reduce the oscillation amplitude}\label{sec.red_ampl}
In this section we investigate two effects responsible for a reduction of the amplitude
of the oscillations of $\bar{J}$ (see main text), as compared to the theoretical prediction
given by Eq.~(6) of the main text.
We first consider the effect of the finite ramp time of the interaction strength, which reduces
the squeezing of the Bogoliubov modes, as compared to an instantaneous quench. 
This effect contributes
to the reduction of the amplitude on the order of 10\%.
We then investigate the reduction of the amplitude induced by the binning of the data with a finite 
resolution in $\tau$. This effect amounts to an additional reduction of the amplitude by 18\%.\\

\subsection{Beyond the instantaneous quench: finite ramp time}\label{SM_ramp_time}
In the experiment, the change of the effective interaction strength is not 
instantaneous: to ensure the adiabatic following of the transverse
motion, we perform a ramp of the transverse oscillation frequency 
during a time $t_r$.
The finite value of  $t_r$
is responsible for a decrease of  the 
induced squeezing of each mode. In the asymptotic limit of very large $t_r$,
the squeezing vanishes
since then, the modes follow adiabatically the modification 
of the interaction strength.
In the following we compute the effect of the ramp on the squeezing
of each mode and we use this result to compute the 
resulting decrease of the oscillation amplitude of $\bar{J}$.

\begin{figure}
  \centerline{\raisebox{1cm}{\scalebox{0.7}{\includegraphics{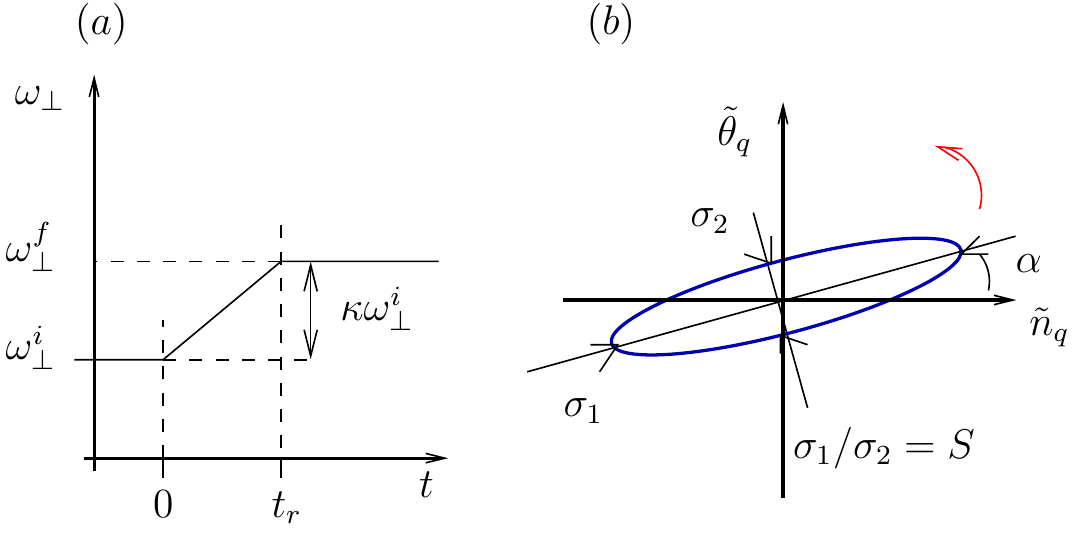}}}}
  \centerline{\includegraphics{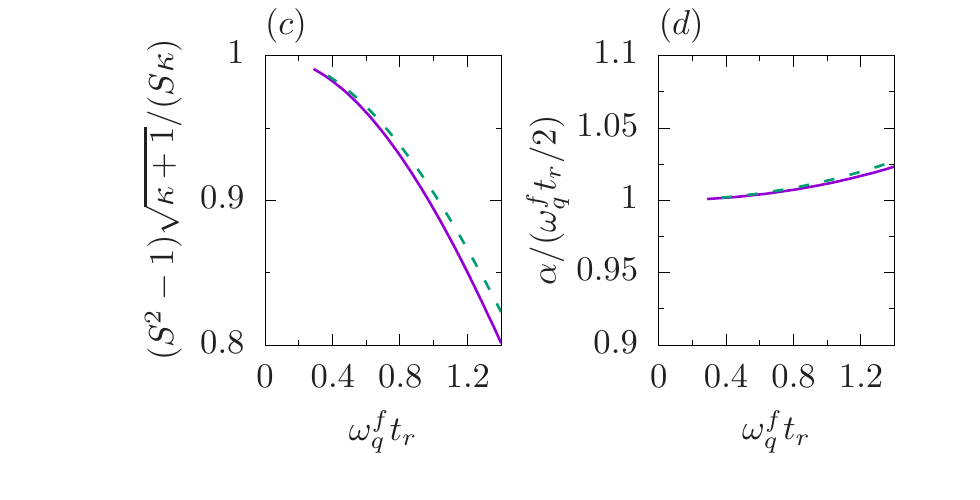}}
\caption{Effect of the interaction strength ramp on the squeezing 
of longitudinal modes. The time sequence is shown in $(a)$. 
An example of  the phase space 
distribution at the end of the ramp
is shown in $(b)$: the $1/\sqrt{e}$ line of the Gaussian 
distribution is plotted. 
  The squeezing factor $S$ is the ratio between the rms widths
along the anti-squeezed and the squeezed directions.
The curved arrow shows the direction of 
rotation under free evolution.
Quantitative results are shown in $(c)$ and $(d)$ 
for a quench strength $\kappa=\omega_\perp^f/\omega_\perp^i-1=2$ (solid lines)
and $\kappa=4$ (dashed lines).
$(c)$ shows 
$\sqrt{\kappa+1}(S^2-1)/(S\kappa)$, 
which gives the amplitude of the resulting breathing 
oscillations normalized to the amplitude for an instantaneous 
quench (see text), versus $\omega_q^f t_r$ where $\omega_q^f$
is the final frequency of the mode. 
The squeezing angle is shown in $(d)$, normalized by $\omega_q^f t_r/2$.
}
\label{fig.effetrampe}
\end{figure}

In order to estimate the effect of the finite ramp time, we 
 will consider a homogeneous gas for simplicity. The Bogoliubov modes
are then described by the Hamiltonian of Eq.~(1) of the main text, namely
\begin{equation}
H_q=A_q n_q^2+B_q \theta_q^2.
\end{equation}
We regard the effect of a 
ramp of $\omega_\perp$ between the time $t=0$ and the time 
$t_r$:  $\omega_\perp$ goes from $\omega_\perp^i$ to
$\omega_\perp^f=(\kappa+1)\omega_\perp^i$, 
 as depicted in Fig.~(\ref{fig.effetrampe}). 
The coefficient $B_q=n_0\hbar^2q^2/(2m)$ is time-independent, while
the coefficient $A_q$ evolves linearly during the ramp ({\it i.e.} during  time interval $0<t<t_r$), since it is proportional to 
$c^2$, itself proportional to $\omega_\perp$. Then,  
the solution of the second order equations describing the evolution of $\theta_q$
and $n_q$ during the ramp 
is given in terms of the Airy functions.
In order to investigate the squeezing, it is natural to 
introduce the reduced variables 
\begin{equation}
\left \{
\begin{array}{l}
\tilde{\theta}_q=\theta_q/\bar{\theta}_q\\
\tilde{n}_q=n_q/\bar{n}_q\\
\end{array}
\right .
\end{equation}
where $\bar{\theta}_q=(A_q(t)/B_q)^{1/4}$ and 
$\bar{n}_q=(B_q/A_q(t))^{1/4}$ are the time-dependent
widths of the ground state. 
For given initial values, the values of  
$\tilde{\theta}_q$ and $\tilde{n}_q$ at the end of the ramp are 
\begin{equation}
\left ( \begin{array}{l}
\tilde{\theta}_q(t_r)\\
\tilde{n}_q(t_r)
\end{array}\right )
=M
\left ( \begin{array}{l}
\tilde{\theta}_q(0)\\
\tilde{n}_q(0)
\end{array}\right )
\label{eq.evolM}
\end{equation}
where the matrix M has the following components:
\begin{widetext}
\begin{equation}
\left \{
\begin{array}{l}
M_{11}=(\kappa+1)^{-1/4}\pi\left ( -B_i(-\delta^{-2/3})A_i'(-(\kappa+1)\delta^{-2/3})
+A_i(-\delta^{-2/3})B_i'(-(\kappa+1)\delta^{-2/3}) \right )\\
M_{22}=(\kappa+1)^{1/4}\pi\left ( B_i'(- \delta^{-2/3})A_i(-(\kappa+1)\delta^{-2/3}) -
A_i'(-\delta^{-2/3})B_i(-\delta^{-2/3}(\kappa+1))\right )\\
M_{21}=(\delta^{-4/3}(\kappa+1))^{1/4}\pi \left ( -B_i(-\delta^{-2/3})A_i(-\delta^{-2/3}(\kappa+1))
+A_i(-\delta^{-2/3})B_i(-\delta^{-2/3}(\kappa+1))\right )\\
M_{12}=(\delta^{-4/3}(\kappa+1))^{-1/4}\pi\left ( B_i'(-\delta^{-2/3})A_i'(-\delta^{-2/3}(\kappa+1))
-A_i'(-\delta^{-2/3})B_i'(-\delta^{-2/3}(\kappa+1))\right )
\end{array}
\right .
\end{equation}
\end{widetext}
Here $A_i,B_i$  are the first and second kind Airy functions 
and $A_i',B_i'$, their derivatives and  $\delta = \kappa/(t_r \omega_q^i )$ the quench speed normalized to 
the initial mode frequency (we recall that the quench strength is 
$\kappa=\omega_\perp^f/\omega_\perp^i-1$).
Under this transformation, the initial isotropic Gaussian distribution transforms 
into a squeezed distribution, {\it i.e.} a Gaussian elliptical distribution with a squeezing
angle $\alpha$ and ratio between the rms width of the two eigenaxes equal to the 
squeezing factor $S$.
In order to find  $\alpha$  and $S$, let us  compute, for any angle
$\beta$, the width 
along the quadrature 
$\tilde{x}_\beta=\cos(\beta)\tilde{\theta}_q +\sin(\beta)\tilde{n}_q$. 
Using  the fact that the initial state is a thermal equilibrium state
fulfilling
$\langle \tilde{\theta}_q^2\rangle_i =\langle \tilde{n}_q^2\rangle_i \equiv V$ and
$\langle \tilde{\theta}_q\tilde{n}_q\rangle_i =0$, and using the 
transformation above,
we find
\begin{multline}
\langle \tilde{x}_\beta^2\rangle =V \lbrace \cos^2(\beta)\left ( M_{11}^2+M_{22}^2 \right )
+\sin^2(\beta)\left ( M_{21}^2 + M_{22}^2\right ) \\
+2\cos(\alpha)\sin(\alpha)\left ( M_{11}M_{21}+M_{22}M_{12}\right ) \rbrace.
\label{eq.effetrampeVar}
\end{multline}
The squeezing angle $\alpha$  is found by imposing $\left . d\langle \tilde{x}_\beta^2\rangle/d\beta\right |_{\beta=\alpha}=0$, which leads to
\begin{equation}
\tan(2\alpha)=-2\frac{M_{11}M_{21}+M_{22}M_{12}}{M_{21}^2+M_{22}^2-M_{11}^2-M_{12}^2}.
\end{equation}
The most squeezed quadrature is $\tilde{x}_{\alpha}$ while $\tilde{x}_{\alpha+\pi/2}$
is the most anti-squeezed quadrature.
The squeezing factor is $S=\sqrt{\langle \tilde{x}_{\alpha}^2\rangle}/\sqrt{\langle \tilde{x}_{\alpha+\pi/2}^2\rangle}$. It also writes 
$S=\langle \tilde{x}_{\alpha}^2\rangle/V$ since the conservation of the phase-space area 
ensures
$\sqrt{\langle \tilde{x}_{\alpha}^2\rangle\langle \tilde{x}_{\alpha+\pi/2}^2\rangle}=V$, 
and it is evaluated  injecting $\beta=\alpha$ in Eq.~(\ref{eq.effetrampeVar}).
Results are shown in Fig.~(\ref{fig.effetrampe}) for quench amplitudes 
$\kappa=2$ and $\kappa=4$ as a function of $\omega_q^f t_r$ where 
$\omega_q^f$ is the final frequency of the mode.
For very slow modes  $\omega_q^f t_r\ll 1$, one recovers the results 
expected for an instantaneous quench~: $\alpha\simeq 0$ and $(S^2-1)\simeq \kappa$.
For modes of larger frequency, the effect of the ramp is to reduce the squeezing 
and also to rotate its axis. 

%\begin{widetext}
%\end{widetext}

The post-quench dynamics results in a 
breathing of the $\tilde{\theta}_q$ quadrature: $\langle \tilde{\theta}_q^2\rangle$
oscillates with an amplitude 
$V(S^2-1)/S$.
Coming back to the variable  ${\theta}_q$, the evolution at times $t>t_r$ writes
\begin{equation}
\langle {\theta}_q^2\rangle(t)=\langle {\theta}_q^2\rangle_i
\frac{\sqrt{\kappa+1}}{S_q}
\left (
1+(S_q^2-1)\sin^2(\omega_q^f (t-t_r) + \alpha_q) 
\right )
\label{eq.squeezingeffetrampe}
\end{equation}
where the indice $q$ in $S$ and $\alpha$ indicates these quantities depend 
on $q$.
As seen in Fig.~(\ref{fig.effetrampe}), the angle $\alpha_q$ is very close to 
$\omega_q^ft_r/2$, for moderate values of $\omega_q^f t_r$. Injecting this 
value into 
Eq.~(\ref{eq.squeezingeffetrampe}), we find that it amounts 
to shifting the time reference 
to $t_r/2$. We perform this shift when analyzing the data, in other terms the reduced variable 
$\tau$ is $\tau=cq(t-t_r/2)$.

Let us now consider the evolution of the density-ripples power spectrum 
$\langle |\tilde{\rho}_q|^2\rangle(t)$. 
For small $q$, 
$\langle |\tilde{\rho}_q|^2\rangle(t)$ is proportional to 
$\langle {\theta}_q|^2\rangle(t)$ such that the evolution of 
$\langle |\tilde{\rho}_q|^2\rangle(t)$ is given by 
Eq.~(\ref{eq.squeezingeffetrampe}). This leads to, 
%=\frac{\langle |\tilde{\rho}_q|^2\rangle(t=\tau/(cq))}{\langle |\tilde{\rho}_q|^2\rangle_i}
\begin{equation}
J(q,\tau)=\frac{\sqrt{\kappa+1}}{S_q}
\left (
1+(S_q^2-1)\sin^2(\tau) 
\right ).
\end{equation}
Let us now investigate the quantity $\bar{J}(\tau)$, defined in the main text 
for experimental data. 
Here we will  assume that
the measurement times are spread over 
$[t_m,t_M]$  and we denote $h(t)dt$ the number of points in the time interval 
$[t,t+dt]$. The $q$ values are assumed to be 
equally spaced, as in the case of a Fast Fourier Transform, 
and only $q$ values in the interval $[q_m,q_M]$ are considered.
 We assume that  $\bar{J}(\tau)$ is obtained by 
binning in $\tau$ the collection of data  
with a bin size $\Delta$ small enough so that,
for all measurement times $t$,  $J(q,\tau)$ is about constant in the 
interval $q \in [\tau/(ct),(\tau+\Delta)/(ct)]$.
Then, one has
\begin{equation}
\bar{J}(\tau)=\frac{1}{\int h(t)dt \Delta/(ct) }
 \int h(t)dt J(q=\tau/(ct),\tau)  \Delta/(ct),
 \end{equation}
where the integrals are evaluated between $t_{1}$ and $t_{2}$, where 
$t_1=\rm{Max}(t_m,\tau/(cq_M))$ and 
$t_2=\rm{Min}(t_M,\tau/(cq_m))$.
Typically, in the experiment  small times are sampled more densely than 
large times. Taking $h$ proportional to $1/t$, we obtain
\begin{equation}
\bar{J}(\tau) = \frac{1}{\int dt/t^2}\int dt \frac{J(q= \tau/(ct),\tau)}{t^2}=\frac{\int dq J(q,\tau)}{q_2(\tau)-q_1(\tau)},
\end{equation}
where $q_1=\rm{max}(\tau/(c)t_M,q_m)$ and 
$q_2=\rm{min}(\tau/(c)t_m,q_M)$.

\begin{figure}
\centerline{\includegraphics{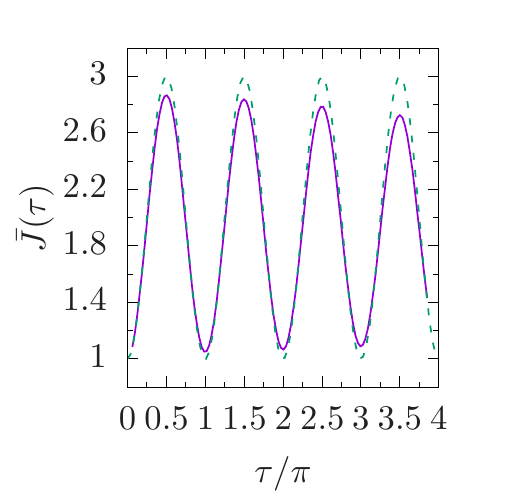}}
\caption{Effect of the finite ramp time of the interaction strength
for a homogeneous gas. 
The expected behavior of $\bar{J}$ (solid line) is compared to the 
case of an instantaneous ramp (dashed line). 
Here we consider a gas of Rubidium atoms at conditions close to 
the experimental ones. More precisely, 
the linear density is $n_0=630$~atoms per 
$\mu$m, the initial transverse oscillation frequency is
$\omega_\perp=2\pi \times 1.5$~kHz, the  quench strength is $\kappa=2$
and the ramp time is $t_r=0.7$~ms. The range of $q$ values 
used to compute $\bar{J}$ is $q\in[0.1,0.5]\mu$m$^{-1}$ and the range 
of measurement times is $t\in [t_r/2,6$ms$]$. }
\label{fig.effetrampeJbar}
\end{figure}

The predicted time evolution of $\bar{J}$ is shown in Fig.~(\ref{fig.effetrampeJbar}) for parameters close
to that of the experimental data shown in the main text. 
The amplitude of the first oscillation is decreased by about 10\%.

\subsection{Finite width of the convolution function used in data processing}
The data shown in the inset of Fig.~(3) of the main text 
correspond to a data set with an exceptionally good signal over noise.
In general, the spread of the data points corresponding to a given value of 
$\tau$ (and thus corresponding to different times  $t$ and wavevectors $q$)
is as large as about 50\%.
In such conditions, a binning of the data as a function of the reduced time $\tau=cqt$ 
with a bin size 
sufficiently large to accommodate many data points    
is required in order to 
increase the signal over noise. As describe in the main text, we use a ``smooth'' binning:
  we compute the weighted average of the data, $\bar{J}$,  with a Gaussian cost function of
  rms width 
$\Delta$.
For a very dense data set, we can define the local average value 
$\tilde{J}(\tau)=\sum_{i,\tau_i\in[\tau,\tau+d\tau]}J_i/d\tau$, where the sum is done on the data set
and $d\tau$ is much smaller than $\Delta$. 
Then $\bar{J}$ corresponds to the
convolution of $\tilde{J}$ with a convolution width $\Delta$.
This convolution  reduces the amplitude of the oscillations.
To estimate this amplitude reduction, let us 
disregard the small damping of the oscillations coming from the cloud inhomogeneity (see section 3)
and thus consider data which would follow
the oscillatory behavior $\tilde{J}=A \sin^2(\tau)$.
 The smoothing $\bar J(\tau) =
\int_{-\infty}^{\infty} \!d\tau' \tilde{J}(\tau') \, e^{-\left(\tau' - \tau
  \right)^2/(2\Delta^2)}/(\sqrt{2 \pi \Delta^2})$ reduces the 
amplitude to $A' = A e^{-2\Delta^2}$. For $\Delta= 0.1\pi$, as used for 
the data analysis shown in the main text, the amplitude is reduced by 18\%.

\bibliographystyle{unsrt}
\bibliography{article-quenchg}

\begin{thebibliography}{10}

\bibitem{polkovnikov_colloquium:_2011}
Anatoli Polkovnikov, Krishnendu Sengupta, Alessandro Silva, and Mukund
  Vengalattore.
\newblock Colloquium: {Nonequilibrium} dynamics of closed interacting quantum
  systems.
\newblock {\em Rev. Mod. Phys.}, 83(3):863--883, August 2011.

\bibitem{Note1}
See~\cite {mitra_quantum_2017} and references therein.

\bibitem{trotzky_probing_2012}
S.~Trotzky, Y.-A. Chen, A.~Flesch, I.~P. McCulloch, U.~Schollwöck, J.~Eisert,
  and I.~Bloch.
\newblock Probing the relaxation towards equilibrium in an isolated strongly
  correlated one-dimensional {Bose} gas.
\newblock {\em Nature Physics}, 8(4):nphys2232, February 2012.

\bibitem{cheneau_light-cone-like_2012}
Marc Cheneau, Peter Barmettler, Dario Poletti, Manuel Endres, Peter Schauß,
  Takeshi Fukuhara, Christian Gross, Immanuel Bloch, Corinna Kollath, and
  Stefan Kuhr.
\newblock Light-cone-like spreading of correlations in a quantum many-body
  system.
\newblock {\em Nature}, 481(7382):484--487, January 2012.

\bibitem{hung_cosmology_2013}
Chen-Lung Hung, Victor Gurarie, and Cheng Chin.
\newblock From {Cosmology} to {Cold} {Atoms}: {Observation} of {Sakharov}
  {Oscillations} in a {Quenched} {Atomic} {Superfluid}.
\newblock {\em Science}, 341(6151):1213--1215, September 2013.

\bibitem{langen_double_2017}
Tim Langen, Thomas Schweigler, Eugene Demler, and Jörg Schmiedmayer.
\newblock Double light-cone dynamics establish thermal states in integrable 1d
  {Bose} gases.
\newblock {\em arXiv:1709.05994 [cond-mat, physics:quant-ph]}, September 2017.
\newblock arXiv: 1709.05994.

\bibitem{jaskula_acoustic_2012}
J.-C. Jaskula, G.~B. Partridge, M.~Bonneau, R.~Lopes, J.~Ruaudel, D.~Boiron,
  and C.~I. Westbrook.
\newblock Acoustic {Analog} to the {Dynamical} {Casimir} {Effect} in a
  {Bose}-{Einstein} {Condensate}.
\newblock {\em Phys. Rev. Lett.}, 109(22):220401, November 2012.

\bibitem{de_nardis_solution_2014}
Jacopo De~Nardis, Bram Wouters, Michael Brockmann, and Jean-Sébastien Caux.
\newblock Solution for an interaction quench in the {Lieb}-{Liniger} {Bose}
  gas.
\newblock {\em Phys. Rev. A}, 89(3):033601, March 2014.

\bibitem{calabrese_interaction_2014}
Pasquale Calabrese and Pierre~Le Doussal.
\newblock Interaction quench in a {Lieb}–{Liniger} model and the {KPZ}
  equation with flat initial conditions.
\newblock {\em J. Stat. Mech.}, 2014(5):P05004, 2014.

\bibitem{cazalilla_quantum_2016}
M.~A. Cazalilla and Ming-Chiang Chung.
\newblock Quantum quenches in the {Luttinger} model and its close relatives.
\newblock {\em J. Stat. Mech.}, 2016(6):064004, 2016.

\bibitem{swislocki_quantum_2016}
Tomasz Świsłocki and Piotr Deuar.
\newblock Quantum fluctuation effects on the quench dynamics of thermal
  quasicondensates.
\newblock {\em J. Phys. B: At. Mol. Opt. Phys.}, 49(14):145303, 2016.

\bibitem{rauer_recurrences_2017}
Bernhard Rauer, Sebastian Erne, Thomas Schweigler, Federica Cataldini,
  Mohammadamin Tajik, and Jörg Schmiedmayer.
\newblock Recurrences in an isolated quantum many-body system.
\newblock {\em arXiv:1705.08231 [cond-mat, physics:quant-ph]}, May 2017.
\newblock arXiv: 1705.08231.

\bibitem{mora_extension_2003}
Christophe Mora and Yvan Castin.
\newblock Extension of {Bogoliubov} theory to quasicondensates.
\newblock {\em Phys. Rev. A}, 67(5):053615, May 2003.

\bibitem{schemmer_monte_2017}
M.~Schemmer, A.~Johnson, R.~Photopoulos, and I.~Bouchoule.
\newblock Monte {Carlo} wave-function description of losses in a
  one-dimensional {Bose} gas and cooling to the ground state by quantum
  feedback.
\newblock {\em Phys. Rev. A}, 95(4):043641, April 2017.

\bibitem{Note2}
For each positive $q$ value, one has 2 Fourier components: $\hat
  {n}_{q,c}=\sqrt {2/L}\int dz n(z)\cos (qz)$ and $\hat {n}_{q,s}=\sqrt
  {2/L}\int dz n(z)\sin (qz)$, with similar expressions for $\theta $. We omit
  the subscript $c$ or $s$ in the text for simplicity.

\bibitem{Note3}
For quasi-1D gases the hydrodynamic condition is replaced by $\omega _q \ll
  \omega _{\perp }$.

\bibitem{Note4}
The phase space area is preserved, one quadrature being squeezed, while the
  other is anti-squeezed.

\bibitem{Note5}
For the $q$ values considered, $\hbar \omega _q \ll k_B T$ and the
  Raighley-Jeans approximation holds.

\bibitem{Note6}
In~\cite {hung_cosmology_2013}, the evolution of density fluctuations has
  however been investigated for a 2D gas.

\bibitem{Note7}
Isolating the contribution of individual modes to the function $g_1(z)$
  requires looking at the Fourier transform of $\protect \qopname \relax
  o{ln}(g_1(z))$, which requires large detection dynamics.

\bibitem{imambekov_density_2009}
A.~Imambekov, I.~E. Mazets, D.~S. Petrov, V.~Gritsev, S.~Manz, S.~Hofferberth,
  T.~Schumm, E.~Demler, and J.~Schmiedmayer.
\newblock Density ripples in expanding low-dimensional gases as a probe of
  correlations.
\newblock {\em Phys. Rev. A}, 80(3):033604, September 2009.

\bibitem{dettmer_observation_2001}
S.~Dettmer, D.~Hellweg, P.~Ryytty, J.~J. Arlt, W.~Ertmer, K.~Sengstock, D.~S.
  Petrov, G.~V. Shlyapnikov, H.~Kreutzmann, L.~Santos, and M.~Lewenstein.
\newblock Observation of {Phase} {Fluctuations} in {Elongated}
  {Bose}-{Einstein} {Condensates}.
\newblock {\em Phys. Rev. Lett.}, 87(16):160406, October 2001.

\bibitem{manz_two-point_2010}
S.~Manz, R.~Bücker, T.~Betz, Ch. Koller, S.~Hofferberth, I.~E. Mazets,
  A.~Imambekov, E.~Demler, A.~Perrin, J.~Schmiedmayer, and T.~Schumm.
\newblock Two-point density correlations of quasicondensates in free expansion.
\newblock {\em Phys. Rev. A}, 81(3):031610, March 2010.

\bibitem{rauer_cooling_2016}
B.~Rauer, P.~Grišins, I. E. Mazets, T.~Schweigler, W.~Rohringer, R.~Geiger,
  T.~Langen, and J.~Schmiedmayer.
\newblock Cooling of a {One}-{Dimensional} {Bose} {Gas}.
\newblock {\em Phys. Rev. Lett.}, 116(3):030402, January 2016.

\bibitem{Note8}
For consistency we rederive this expression (first established in~\cite
  {imambekov_density_2009}) see Appendix~\ref {SM_Density_ripples_hom},\ref
  {SM_Density_ripples_LDA}.

\bibitem{Note9}
In Eq.~(\ref {eq.drasymp}), $\langle \theta _q^2 \rangle =(\langle \theta
  _{q,c}^2 \rangle +\langle \theta _{q,s}^2 \rangle )/2$ where $\theta _{q,c}$
  and $\theta _{q,s}$ are the cosine and sine Fourier components, which fulfill
  $\langle \theta _{q,c}^2\rangle =\langle \theta _{q,s}^2\rangle $ for
  translationally invariant systems.

\bibitem{Note10}
Validity of LDA is established in Appendix~\ref {sec.beyondLDA}.

\bibitem{Note11}
The experiment is described in more detail in~\cite {jacqmin_momentum_2012}.

\bibitem{kheruntsyan_pair_2003}
K.~V. Kheruntsyan, D.~M. Gangardt, P.~D. Drummond, and G.~V. Shlyapnikov.
\newblock Pair {Correlations} in a {Finite}-{Temperature} 1d {Bose} {Gas}.
\newblock {\em Phys. Rev. Lett.}, 91(4):040403, July 2003.

\bibitem{fuchs_hydrodynamic_2003}
J.~N. Fuchs, X.~Leyronas, and R.~Combescot.
\newblock Hydrodynamic modes of a one-dimensional trapped {Bose} gas.
\newblock {\em Phys. Rev. A}, 68(4):043610, October 2003.

\bibitem{Note12}
the box $L$ is chosen to be about twice the size of the cloud.

\bibitem{Note13}
The transverse size of the cloud after the time-of-flight is comparable to the
  depth of focus of the imaging system and depends on the transverse
  confinement. We thus expect slightly different optical resolutions, $\sigma
  _i$ and $\sigma _f$ for data taken before and after the quench respectively.
  We correct for this effect to make quantitative comparison of data taken
  before and after the quench.

\bibitem{Note14}
The values of resolution obtained by such fits are close to the expected values
  if one takes into account the depth of field of our imaging system and the
  fact that, after the the expansion time $t_f$ the cloud explores about 50$\mu
  $m along the imaging direction.

\bibitem{jacqmin_sub-poissonian_2011}
Thibaut Jacqmin, Julien Armijo, Tarik Berrada, Karen~V. Kheruntsyan, and
  Isabelle Bouchoule.
\newblock Sub-{Poissonian} {Fluctuations} in a 1d {Bose} {Gas}: {From} the
  {Quantum} {Quasicondensate} to the {Strongly} {Interacting} {Regime}.
\newblock {\em Phys. Rev. Lett.}, 106(23):230405, June 2011.

\bibitem{Note15}
At a time $t = t^{g_1}_{\protect \rm th} $ the $g_{1}$ function has reached the
  thermal value $e^{-|z|/l_c^f}$ for all $z < 2l_c^f$, the deviation from this
  thermal state being restricted to long distances where $g_{1}(z)<e^{-2}
  \approx 10\%$.

\bibitem{johnson_long-lived_2017}
A.~Johnson, S.~S. Szigeti, M.~Schemmer, and I.~Bouchoule.
\newblock Long-lived nonthermal states realized by atom losses in
  one-dimensional quasicondensates.
\newblock {\em Phys. Rev. A}, 96(1):013623, July 2017.

\bibitem{Note16}
We corrected the formula published in~\cite {imambekov_density_2009}.

\bibitem{hellweg_phase_2001}
D.~Hellweg, S.~Dettmer, P.~Ryytty, J.~J. Arlt, W.~Ertmer, K.~Sengstock, D.~S.
  Petrov, G.~V. Shlyapnikov, H.~Kreutzmann, L.~Santos, and M.~Lewenstein.
\newblock Phase fluctuations in {Bose}–{Einstein} condensates.
\newblock {\em Appl Phys B}, 73(8):781--789, December 2001.

\bibitem{Note17}
This term is due to the approximation made when going from Eq.~(\ref
  {eq.powersectrumdef}) to Eq.~(\ref {eq.pstf}), which is valid only for $q$
  values larger than the inverse of the cloud size.

\bibitem{Note18}
Since the Hamiltonian of interest is quadratic in $\theta $, the distribution
  of $\theta $ is Gaussian at thermal equilibrium. The squeezing of each
  collective mode produced by the interaction quench preserves the Gaussian
  nature of $\theta $.

\bibitem{olshanii_atomic_1998}
M.~Olshanii.
\newblock Atomic {Scattering} in the {Presence} of an {External} {Confinement}
  and a {Gas} of {Impenetrable} {Bosons}.
\newblock {\em Phys. Rev. Lett.}, 81(5):938--941, August 1998.

\bibitem{mitra_quantum_2017}
Aditi Mitra.
\newblock Quantum quench dynamics.
\newblock {\em arXiv:1703.09740 [cond-mat]}, March 2017.
\newblock arXiv: 1703.09740.

\bibitem{jacqmin_momentum_2012}
Thibaut Jacqmin, Bess Fang, Tarik Berrada, Tommaso Roscilde, and Isabelle
  Bouchoule.
\newblock Momentum distribution of one-dimensional {Bose} gases at the
  quasicondensation crossover: {Theoretical} and experimental investigation.
\newblock {\em Phys. Rev. A}, 86(4):043626, October 2012.

\end{thebibliography}

\end{document}